\documentclass[aps, prd, preprint, amsfonts, floatfix]{revtex4}

\usepackage{graphicx}
\usepackage{amsmath,amsfonts}
\usepackage{epsfig,color}

\begin{document}
\newcommand{\be}{\begin{eqnarray}}
\newcommand{\ee}{\end{eqnarray}}
\newcommand\del{\partial}
\newcommand\nn{\nonumber}
\newcommand{\Tr}{{\rm Tr}}
\newcommand{\Str}{{\rm STr}}
\newcommand{\mat}{\left ( \begin{array}{cc}}
\newcommand{\emat}{\end{array} \right )}
\newcommand{\vect}{\left ( \begin{array}{c}}
\newcommand{\evect}{\end{array} \right )}
\newcommand{\tr}{{\rm Tr}}
\newcommand{\hm}{\hat m}
\newcommand{\ha}{\hat a}
\newcommand{\hz}{\hat z}
\newcommand{\hx}{\hat x}
\newcommand{\tm}{\tilde{m}}
\newcommand{\ta}{\tilde{a}}
\newcommand{\tz}{\tilde{z}}
\newcommand{\tx}{\tilde{x}}
\definecolor{red}{rgb}{1.00, 0.00, 0.00}
\newcommand{\rd}{\color{red}}
\definecolor{blue}{rgb}{0.00, 0.00, 1.00}
\definecolor{green}{rgb}{0.10, 1.00, .10}
\newcommand{\blu}{\color{blue}}
\newcommand{\green}{\color{green}}



\title{Spectrum of the Wilson Dirac Operator at Finite Lattice Spacings}
\author{G. Akemann}
\affiliation{Department of Mathematical Sciences $\&$ BURSt Research Centre\\
Brunel University West London. Uxbridge UB8 3PH, United Kingdom }
\author{P.H. Damgaard}
\affiliation{The Niels Bohr International Academy and Discovery Center, 
The Niels Bohr
    Institute, Blegdamsvej 17, DK-2100, Copenhagen
  {\O}, Denmark} 
\author{K. Splittorff}
\affiliation{The Niels Bohr Institute, Blegdamsvej 17, DK-2100, Copenhagen
  {\O}, Denmark} 
\author{J.J.M. Verbaarschot}
\affiliation{Department of Physics and Astronomy, SUNY, Stony Brook,
 New York 11794, USA}

\date   {\today}
\begin  {abstract}
We consider the effect of discretization errors on 
the microscopic spectrum of the Wilson Dirac operator
using both chiral Perturbation Theory and chiral Random Matrix
Theory. A graded chiral Lagrangian is used 
to evaluate  the microscopic spectral density of the 
Hermitian Wilson Dirac operator as well as the distribution 
of the chirality over the
real eigenvalues of the Wilson Dirac operator. 
It is shown that a chiral Random Matrix Theory for the Wilson 
Dirac operator reproduces the leading zero-momentum terms of Wilson
chiral Perturbation Theory.
All results are obtained for  fixed index of the
Wilson Dirac operator. The low-energy constants of Wilson chiral Perturbation 
theory are shown to be constrained by the Hermiticity properties 
of the Wilson Dirac operator. 
\end{abstract}
\maketitle

\section{Introduction}

In the continuum, low-lying spectra of the Dirac operator
for theories with spontaneous chiral symmetry breaking have
two equivalent descriptions. One is in terms of chiral Random
Matrix Theory \cite{SV,RMT,RMT1,RMT2}, and 
the other one is in terms of a chiral Lagrangian \cite{LS,DOTV,SplitV}. 
It is by now well-established how the two formulations are in
one-to-one correspondence  \cite{DOTV,TV,Basile}, and that this 
is valid for 
all spectral correlation functions of the Dirac operator. Even 
single eigenvalue distributions can be derived in both 
formalisms, and have been shown to be equivalent \cite{AD}. The equivalence
is valid to leading order in a chiral counting scheme known as the
$\epsilon$-regime or, in Random Matrix Theory terminology, 
the microscopic domain. 

Apart from its conceptual value, the theory of low-lying Dirac operator 
spectra has been of  quite practical use in lattice gauge theory. In fact,
it serves multiple purposes,  all relying on a QCD partition function that is
formulated from the outset  
at finite four-volume $V$: (1) it can be used to establish 
spontaneous chiral symmetry breaking in a clean way, (2) it provides
precise non-perturbative analytical predictions that can be used to test
the chiral limit, and (3) it allows for a determination of low-energy
constants by means of finite-volume scaling. So far these uses have
been limited by the fact that violations due to finite lattice
spacings, $a$, have been ignored. Such lattice artifacts evidently
depend on the particular lattice discretization chosen.
Following Symanzik's program, corrections due to finite lattice spacings
can, when they are sufficiently small, 
be analyzed in a continuum field theory language through the introduction
of higher-dimensional operators in the Lagrangian. The corrections
to the chiral Lagrangian that arise up to and including order $a^2$-effects
for Wilson fermions have been analyzed in a series of papers 
\cite{SharpeSingleton,RS,BRS,Aoki-spec,Aoki-pion-mass,Golterman:2005ie}.
For  comprehensive 
reviews of effective field theory methods
at finite lattice spacings, see, $e.g.$, ref. \cite{sharpe-nara,Golterman}.

It is then an obvious problem to investigate the effect of
lattice-induced scaling violations on the spectral
properties of the Wilson Dirac operator $D_W$ \cite{Sharpe2006}.
In a recent Letter \cite{DSV}, three of the present authors
have taken up this issue and shown how the microscopic scaling
regime can be phrased
both in terms of the (graded) chiral Lagrangian and a modified
chiral Random Matrix Theory that incorporates finite lattice spacing
effects of Wilson fermions.

Although the Wilson Dirac operator is not Hermitian, the operator
$D_5 \equiv \gamma_5(D_W+m)$ is Hermitian, and much more convenient
to work with in lattice QCD simulations. In this paper we analyze the
microscopic spectrum of this operator. Contrary to the lattice
QCD Dirac operator at $a= 0$, its eigenvalues $ \lambda_k^5(m)$ are
not paired and are nontrivial functions of the quark mass.

There is a deep relation between the topology of gauge field configurations
and the spectrum of the Dirac operator. Not only is the number zero
eigenvalues equal to the difference of the number of right handed and
left handed zero modes, because of level repulsion the Dirac spectrum
near zero is affected in a universal way by the topological charge. 
 As can be shown from spectral flow 
arguments (see, $e.g.$ \cite{SmitVink,Itoh,neuberger,Edwards,Hernandez,Gattringer,Hasenfratz}),
the eigenvalues corresponding to the chiral modes at zero lattice spacing
(when the Dirac operator is antihermitian)
correspond to exactly real eigenvalues at nonzero lattice spacing. Additional
pairs of real modes  appear  for increasing lattice spacing. However, the
number of spectral flow lines, $\lambda_k^5(m)$, with an odd number of
real zeros remains the same. We will identify the number of such flow 
lines as the index of the Dirac operator and study Dirac spectra for
a fixed index. In the continuum limit this index, by the Atiyah-Singer
index theorem becomes equal to the topological charge of the gauge
field configurations. In Random Matrix Theory, this index  
is determined by the block structure of the Dirac matrix. 
All results in this paper are derived for fixed index of the 
Dirac operator. It is of course also
possible to sum over all sectors with a given index.

In this paper we elaborate on and provide detailed derivations of results 
announced in the Letter \cite{DSV} and the proceedings \cite{latNf0}.  
One of the simplifying 
features of that paper was that double-trace terms in the
chiral Lagrangian were ignored (which can be justified 
based on large-$N_c$ arguments \cite{KL}).
 Here we compute their contribution 
to the microscopic spectrum directly from the chiral Lagrangian. 
We also show how double-trace terms can be included in the Wilson chiral
Random Matrix Theory. We study in detail the $a\to0$ and
four-volume $V\to\infty$ limits of the analytical results.  Furthermore, 
it is shown that the low-energy constants of
Wilson chiral Perturbation Theory are constrained by a QCD inequality
which follows from the Hermiticity properties of the Wilson Dirac 
operator. This constraint coincides precisely with the requirement
of preservation of the Hermiticity properties of the Wilson Dirac
operator. The constraints found are consistent with the existence of
an Aoki phase.  
One important message from the calculations presented here
is that the values of the  low-energy constants of Wilson Chiral
Perturbation Theory can be accurately determined by 
matching the predictions for the eigenvalue distributions to 
lattice data.  
Furthermore, our results describe analytically the eigenvalues that may 
cause numerical instabilities \cite{Luscher} when approaching 
the chiral limit at finite lattice spacing with Wilson fermions.

 The analysis of the Letter \cite{DSV} has been extended in various other
directions. We have obtained the distribution of the chirality over the
real eigenvalues of the Wilson Dirac operator which is a lower bound
for the distribution of the real eigenvalues. We also have obtained
an upper bound which converges to the lower bound for small $a$.
In the same limit the expressions for the microscopic spectral density
of $D_5$ simplify and can be generalized to a nonzero number of flavors.
We discuss the distribution of the tail states in the gap
as well as a comparison with the scaling properties of such
states as found in lattice simulations. 
We also perform a saddle point analysis of our analytical result and obtain
a simple explicit expression for the gap of the Dirac spectrum.

The paper is organized as follows. 
After a brief review of relevant 
properties of the Wilson Dirac operator given in section 
\ref{genwil} we define Wilson chiral Perturbation Theory at a
fixed index of the Wilson Dirac Operator in section \ref{sec:WCPT}.
The derivation of the microscopic spectrum is given in section 
\ref{sec:eff}. It is followed by a discussion of various limits 
in section \ref{sec:small_a}. A chiral Random Matrix Theory for Wilson
fermions is shown to reproduce Wilson chiral Perturbation Theory in 
the microscopic limit in section \ref{sec:WRMT}. Finally,
before concluding, we discuss the relation between the 
low energy constants of the Wilson chiral Perturbation Theory 
and the Hermiticity properties of the Wilson Dirac operator.
Some technical details are referred to \ref{app:A}, \ref{app:B} and
\ref{app:C}.

\section{Eigenvalues of the Wilson Dirac Operator}
\label{genwil}

We start with a discussion of general properties  of the Wilson Dirac
operator. Some of these results have been discussed extensively in the literature 
(see, $e.g.$, refs. 
\cite{Itoh,neuberger,Edwards,Hernandez,Gattringer,Sharpe2006}), 
but are included here to make this paper self-contained.

The Wilson-Dirac operator will be denoted by $D = D_W + m$. The lowest
order $a$ correction was introduced by Wilson, 
\be
D_W = \frac{1}{2}\gamma_{\mu}(\nabla_{\mu} + \nabla^*_{\mu})
-\frac{1}{2}a\nabla^*_{\mu}\nabla_{\mu}.
\ee 
It is written in terms
of forward ($\nabla_{\mu}$) and backward ($\nabla^*_{\mu}$) covariant 
derivatives,
and $m$ is the quark mass.

The Wilson Dirac operator, $D_W$, is not anti-Hermitian at non-zero lattice
spacing.  
It retains only $\gamma_5$-Hermiticity: 
\be
D_W^\dagger = \gamma_5 D_W \gamma_5. 
\label{g5Herm}
\ee
The eigenvalues, $\lambda^W_k$, of $D_W$ are then 
either real or they occur in complex conjugate pairs. 
The $\gamma_5$-Hermiticity of $D_W$ implies that the operator
\be
D_5 \equiv \gamma_5 (D_W +  m)
\ee
is Hermitian. Since at non-zero lattice spacings $a$ 
the axial symmetry is lost ($\{D_W, \gamma_5 \}\ne 0$), 
the eigenvalues of $D_5$ do not occur in pairs of opposite sign.
 
The eigenvalues, $\lambda_k^5(m)$, of $D_5$ are  nontrivial functions 
of $m$. If
$ \lambda_k^5(m_c) = 0$, then
\be
\gamma_5(D_W+m_c) \phi = 0 \quad \Rightarrow \quad D_W\phi = -m_c \phi,
\ee
so that the real eigenvalues of $D_W$ can be obtained from the zeros
of $\lambda_k^5(m)$ with the corresponding eigenfunctions of $D_5$ and
$D_W$ beeing identical at $m=m_c$. 
If we act with $D_5$ on the normalized eigenfunctions
$\phi_j\equiv|j\rangle$ of $D_W$ we obtain
\be
D_5 \phi_j = \gamma_5(D_W+m ) \phi_j = (-m_c+m) \gamma_5 \phi_j.
\ee
For $m$ close to $m_c$ 
\be 
\gamma_5 \phi_j = \langle j |\gamma_5 | j \rangle\phi_j  +\delta \phi,
\ee
with $\langle \delta \phi | j \rangle = 0$ and $ \delta \phi \sim O(m-m_c)$.
Therefore to $O(m-m_c)$ 
\be
\lambda_j^5(m) = (m -m_c) \langle j |\gamma_5 | j \rangle. 
\ee
We thus find \cite{Itoh}
\be
\frac{d\lambda_j^5}{dm} = \langle j |\gamma_5 | j \rangle +O(m-m_c).
\ee
When evaluated at $m=m_c$, ie. at $\lambda_j^5=0$, the slope is
thus exactly given by the chirality 
\be
\frac{d\lambda_j^5}{dm}\Big|_{m=m_c} = \langle j |\gamma_5 | j \rangle.
\label{slope}
\ee
Furthermore, it can be shown \cite{Itoh} that the chirality of the
eigenfunctions 
\be 
\chi_j \equiv \langle j| \gamma_5| j \rangle
\ee
vanishes for complex eigenvalues of $D_W$ but is generally nonzero for 
real eigenvalues.

To compute the spectral density $\rho_5$  of $D_5$ 
we introduce a source $z$ that couples to $\bar{\psi}\gamma^5\psi$. 
The operator entering in the fermion determinant is then given by
\be
\gamma_5(D_W + m)+z = D_5+z,
\ee
and the  resolvent of $D_5$ is defined by
\be\label{Gdef}
G(z) \equiv  \left \langle   \tr \left(\frac{1}{D_5+z -i\epsilon}\right)\right\rangle =
\left \langle \sum_k \frac 1{\lambda^5_k +z-i\epsilon}\right \rangle,
\ee
where $\lambda^5_k$ are the eigenvalues of $D_5$.
The density of eigenvalues of $D_5$ then follows from
\be
\rho_5(\lambda^5,m;a) = \left \langle \sum_k \delta(\lambda^5_k-\lambda^5) \right \rangle = \frac{1}{\pi}{\rm Im}[G(-\lambda^5)]_{\epsilon\to0}.
\label{rho5def}
 \ee

The resolvent  (\ref{Gdef}) is  a partially
quenched chiral condensate. We will compute it by means of the graded
technique, where the unphysical 'valence' determinant is canceled by
an inverse determinant after differentiation with respect to the
source \footnote{To compute the spectrum of $D_W$ in the complex plane, a  
different generating function that contains in addition a complex
conjugate valence fermion determinant is needed.}.
In order to be able to write the inverse determinant as a bosonic integral,
it is essential that an infinitesimal increment is added to a Hermitian 
operator \cite{Golterman:2005ie}.

We also consider the chiral condensate corresponding to the regularization 
introduced in Eq.~(\ref{Gdef}) 
\be\label{Sigmadef}
\Sigma(m)\equiv\left\langle{\rm Tr}\frac 1{D_W+m-i\epsilon\gamma_5}\right\rangle.
\ee
Care has to be taken in order to relate $\Sigma(m)$ to the spectrum of 
$D_W$. The discontinuity of $\Sigma(m)$ across the real axis is
given by
\be
\rho_\chi(m) &\equiv&
\frac 1{2\pi i} 
\left\langle{\rm Tr} \left[\frac 1 {(D_W +m) -i\epsilon\gamma_5}
-\frac 1 { (D_W +m) +i\epsilon\gamma_5}\right]_{\epsilon\to0}\right\rangle
\nn \\
&=&  \frac 1{2\pi i} 
\left\langle
{\rm Tr} \left[\frac {\gamma_5} {\gamma_5 (D_W +m)-i\epsilon}
-\frac  {\gamma_5} {\gamma_5 (D_W +m)+i\epsilon}\right]_{\epsilon\to0}
\right\rangle.
\label{chidef}
\ee
This expression can be rewritten in terms of eigenvalues and the
normalized eigenfunctions $|k\rangle$ of $D_5$ (recall that for
$\lambda^5_k=0$ the eigenfunctions of $D_5$ merge with that belonging
to a real mode of $D_W$)
\be
\rho_\chi(m) &=& \frac 1\pi \left\langle
\sum_k \frac {\epsilon\ \langle k| \gamma_5 | k \rangle}
{(\lambda^5_k(m))^2 +\epsilon^2}\Big|_{\epsilon\to0} \right\rangle
\nn \\
&=&   \left\langle
\sum_k \delta(\lambda_k^5(m))\  \langle k| \gamma_5 | k \rangle
\right\rangle
,\nn \\
&=&  \left\langle\sum_{\lambda_k^W\in {\mathbb R}} 
\frac{1}{|d\lambda_k^5/dm|}\delta(\lambda_k^W +m) \langle k| \gamma_5 | k \rangle\right\rangle,
\ee
where $\lambda_k^W$ are the eigenvalues of $D_W$.  
Using Eq. (\ref{slope}) this can be written as 
\be
\label{distOFchi}
\rho_\chi(\lambda^W) = \left\langle\sum_{\lambda_k^W\in {\mathbb R}}\delta(\lambda_k^W +\lambda^W) \, {\rm sign}(\langle k| \gamma_5 | k \rangle)\right\rangle.
\ee
This shows that $\rho_\chi$ is the distribution of the chiralities
over the real eigenvalues of $D_W$. 
 Integrating this expression over $\lambda^W$ we obtain 
\be
\int \rho_\chi(\lambda^W) d\lambda^W &=& \left\langle\sum_{\lambda_k^W
 \in {\mathbb R}} {\rm sign}(\langle k| \gamma_5 | k \rangle)\right\rangle.
\label{index}
\ee
This is the average index of the Dirac operator. The index for a given
gauge field configuration is defined by 
\be
\label{defIndex}
\nu = \sum_{\lambda_k^W \in {\mathbb R}} {\rm sign} (\langle k
|\gamma_5|k\rangle).
\ee

Since the real modes 
of $D_W$ correspond  to the vanishing eigenvalues of $D_5$ we 
also consider $\rho_5(\lambda^5=0,m) $.
A similar calculation as above results in
\be
\rho_5 (\lambda^5=0,m;a) =  \left\langle\sum_{\lambda_k^W
 \in {\mathbb R} }\frac {\delta(\lambda_k^W +m)}{|\langle k| \gamma_5 | k
  \rangle|}\right\rangle. 
\ee
Because $|\langle k| \gamma_5 | k\rangle| \le 1$ we have the inequality
(also valid for fixed index)
\be\label{rhoINEQ}
\rho_\chi(\lambda^W) \le \rho_{\rm real}(\lambda^W) \le 
\rho_5(\lambda^5=0,m=\lambda^W;a),
\ee
where 
\be
\rho_{\rm real}(\lambda^W) \equiv \left\langle\sum_{\lambda_k^W
 \in {\mathbb R} }\delta(\lambda_k^W +\lambda^W)\right\rangle.
\ee 
As a special case the average number, $\langle n_{\rm real}\rangle$,
of real eigenvalues of $D_W$ for gauge field configurations with index
$\nu$ is bounded by 
\be
\int d\lambda^W \, \rho^\nu_\chi(\lambda^W)=\nu 
\ \leq  \ \langle n_{\rm real}\rangle \ \leq \ \int dm \, \rho^\nu_5(\lambda^5=0,m). 
\ee

\section{Wilson chiral Perturbation Theory}
\label{sec:WCPT}

The operators contributing to Wilson chiral Perturbation Theory 
have been considered in a series of papers \cite{SharpeSingleton,RS,BRS}. 
It is constructed in
terms of a double expansion in both the usual parameters
of continuum chiral Perturbation Theory {\em and} the lattice spacing $a$.
Depending on which counting scheme one uses, various terms contribute to a
given order \cite{Shindler:2009ri, Bar:2008, Bar:2010}. We concentrate on the
microscopic limit, the $\epsilon$-regime, in which the combinations
\be
m V, \quad z V, \quad a^2 V
\ee
are kept fixed in the infinite-volume limit $V\to\infty$. To leading
order in these quantities, the low energy
partition function for lattice QCD with Wilson fermions 
then reduces to a unitary matrix integral. Up to the infinite-volume
chiral condensate $\Sigma$
and three low-energy constants determined by the discretization 
errors of Wilson lattice QCD, this integral is 
completely determined by symmetry arguments.

Since we are computing quantities up to and including
${\cal O}(a^2)$ effects, it is important that all effects to this
order are contained in the on-shell effective Symanzik action including
${\cal O}(a^2)$ terms. This problem has been analyzed in detail by
Sharpe  for the $p$-counting-scheme \cite{Sharpe2006}. 
Here we reconsider the argument for the $\epsilon$-regime. 
One  correction to relevant continuum operators corresponds to 
$a^2 \bar{\psi}(x)\gamma^5 \nabla_\mu^2\psi(x)$ \cite{Sharpe2006}. In
the $\epsilon$-regime, such terms are suppressed 
by $ 1/\sqrt V $ with respect to the leading
terms and need to be considered only at
higher orders in the expansion. Additional lattice artifacts of
potential ${\cal O}(a^2)$ in
the form of contact terms have been analyzed in ref. \cite{Sharpe2006},
and they are similarly suppressed.

As stressed by Leutwyler and Smilga \cite{LS}, the topological charge
of gauge field configurations
plays an important role in the microscopic spectrum of the 
continuum theory.
The $\nu$ zero modes of the Dirac operator distort the
eigenvalue spectrum of the non-zero modes and lead to distinct
predictions in sectors of fixed $\nu$. 
Likewise it is natural to study the microscopic spectral density of 
the Wilson Dirac operator for fixed index $\nu$.
 At the level of the chiral Lagrangian 
it is {\em a priori} 
far from obvious how to implement the projection onto sectors 
corresponding to a fixed index of $D_W$.
Quite remarkably, 
such a projection can be achieved through a Fourier
transform \cite{DSV}: We decompose the partition function as 
\be
Z_{N_f}(m,z; a)  = \sum_\nu Z_{N_f}^\nu(m,z;a)
\ee
with
\be 
Z_{N_f}^\nu(m,z;a) & = & \int_{U(N_f)} d U \ {\det}^\nu U
 \ e^{V {\cal L}(U)}.
\label{Zfull}
\ee
The Lagrangian ${\cal L}(U)$ is defined by 
\be\label{ZLfull}
{\cal L}(U) & = &
\frac 12 (m+z) \Sigma { \rm Tr} U
+\frac 12 (m-z)\Sigma{ \rm Tr} U^\dagger \\
&& -a^2W_6[{\rm Tr}\left(U+U^\dagger\right)]^2
     - a^2W_7[{\rm Tr}\left(U-U^\dagger\right)]^2
    - a^2 W_8{\rm   Tr}(U^2+{U^\dagger}^2). \nn
\ee
Below we will demonstrate that in the microscopic domain this 
corresponds to an ensemble of gauge field configurations for which
the index of $D_W$ as defined by Eq. (\ref{defIndex}) is equal to  $\nu$.
 The partitioning into sectors of fixed $\nu$ corresponds
at the level of the unitary group integral to writing an $SU(N_f)$
integral as an infinite sum of $U(N_f)$ integrals. 

In the Lagrangian (\ref{ZLfull}), $\Sigma$ is the usual infinite-volume
chiral condensate while $W_6$, $W_7$ 
and $W_8$ are low-energy constants that quantitatively determine the
discretization errors of Wilson fermions. 
The terms proportional to $W_6$ and $W_7$ are believed to be
suppressed in the large $N_c$ limit \cite{KL}, and they were 
not included in ref. \cite{DSV}.  
Here we include the effect of these terms explicitly. To lighten the 
notation, we introduce the scaling variables
\be
\hm \equiv m\Sigma V, 
\quad \hz \equiv z\Sigma V, \quad \ha_i^2 \equiv a^2W_i V
\ee
with $i=6,7,8$.

\section{Graded Generating Function}
\label{sec:eff}

To obtain the distribution, $\rho_\chi^\nu$, of the chirality over
the real eigenvalues of $D_W$ and $\rho_5^\nu$ 
at fixed index $\nu$ we will
use the graded method. The generating function in this case takes the form 
\be
{\cal Z}^\nu_{N_f+1|1}(m,m',z,z';a) =
 \int ({\rm d}A_\mu)_\nu
\det(D_W+m)^{N_f} 
\frac{\det(D_W + m + \gamma_5z)}{\det(D_W +m' +\gamma_5(z'-i\epsilon))} \ 
\mbox{e}^{-S_{\rm YM}},
\ee
and we have two spectral resolvents
\be
\label{sigma2}
\Sigma^{\nu}(m;a)= -\lim_{m'\to m} \frac{d}{dm'} {\cal
  Z}^\nu_{N_f+1|1}(m,m',z=0,z'=0;a), 
\ee
and
\be
\label{Gsusy}
G^\nu(z,m;a)= -\lim_{z'\to z} \frac{d}{dz'} {\cal Z}^\nu_{N_f+1|1}(m,m,z,z';a).
\ee
Note that the $N_f$ physical quark flavors are not coupled to the
source $z$.

The presence of fermionic as well as bosonic quarks in this generating
function gives rise to a graded structure (``supersymmetry'') of the
corresponding chiral Lagrangian
\be
\label{ZSUSY}
Z^\nu_{N_f+1|1}({\cal M},{\cal Z};\ha_i)  & = & \int \hspace{-1.5mm} dU \
{\rm Sdet}(U)^\nu \\
&& \hspace{-2.5cm}\times 
  e^{i\frac{1}{2}{\Str}({\cal M}[U-U^{-1}])
    +i\frac{1}{2}{\Str}({\cal Z}[U+U^{-1}])
    +\ha_6^2[{\Str}(U-U^{-1})]^2
    +\ha_7^2[{\Str}(U+U^{-1})]^2
    +\ha_8^2{\Str}(U^2+U^{-2})}, \nn 
\ee
where ${\cal M}\equiv{\rm diag}(\hm\ldots \hm,\hm,\hm')$ and 
${\cal Z}\equiv{\rm diag}(0\ldots 0,\hz,\hz')$, with ${\rm Im}(\hz')<0$. 
The integration is over the maximum Riemannian graded submanifold of
$Gl(N_f+1|1)$ \cite{DOTV} (see \cite{DOTV,Efetov} for notation and more on
the graded method). For $N_f = 0$ we will use the parametrization
\be
U &=& \left(\begin{array}{cc} e^{i\theta} & 0 \\ 0 & e^{s}
  \end{array}\right)
\exp\left(\begin{array}{cc} 0 & \alpha \\ \beta & 0 \end{array}\right).
\label{U}
\ee

If we could $derive$ this partition function directly from
the microscopic theory we would end up with $convergent$ integrals. Relying
only on symmetry arguments is not sufficient 
to obtain the partition function, and it can  only be
 equivalent to the microscopic partition function if  the
integrations  are convergent.
For fermionic integrals convergence is not an issue, but bosonic integrals
necessarily have to be convergent.

In \cite{DOTV} the graded  partition function was evaluated 
for the anti-Hermitian Dirac operator with positive mass and 
$\ha_i = 0$ and $\hz = 0$. The term relevant for the convergence of the 
integral was $\exp(-\hm\cosh s)$ so that the generating function 
is defined for $\hm$ on the imaginary axis
with a positive real infinitesimal increment, and can be used to
calculate the spectral density localized on the imaginary axis. We could
also have calculated the spectrum of $\gamma_5 D$ by including the source
term $\hz \gamma_5$. This would have resulted in the bosonic integrand
$\exp[-\hm\cosh s -\hz \sinh s]$. The corresponding
integral can be continued analytically to the entire real $\hz$-axis
by shifting the $s$-integration by $ s \to s \pm \pi i/2$ with the sign
determined by the imaginary part of $\hz$. This transformation can be
accomplished by $ U \to iU $ for ${\rm Im}(\hz) < 0$ and      
by $ U \to -iU $ for ${\rm Im}(\hz) > 0$. The expression (\ref{ZSUSY}) is thus
valid for ${\rm Im}(\hz) < 0$.

Next we consider the  partition function with $\ha_i^2 \ne 0$. Let us
first consider 
the case that $\ha_6 = \ha_7 = 0$, and we consider the partition function
as an analytical function of $\ha_8^2$. The partition function (\ref{ZSUSY}) 
is analytic in $\ha_8^2$ for  ${\rm Re}\, \ha_8^2 > 0 $, whereas the
partition function  
of Ref. \cite{DOTV} can only be continued to ${\rm Re}\, \ha_8^2 < 0 $.
Depending on whether we consider the spectrum of $D_W$ or $\gamma_5 D_W$
the  partition function can be analytically continued to a different part of
the complex $\ha_8^2$-plane. Below we will
argue that the physically interesting parameter domain is  ${\rm Re}\,
\ha_8^2 > 0 $, 
which is accessible for the integration contours of Eq. (\ref{ZSUSY}) valid
for the evaluation of 
the spectrum of $D_5$.
Given this parametrization, the partition
function is defined in only a subset of the complex $\ha_6$ and $\ha_7$ plane.
For real constants the condition is that
$\ha_8^2 -\ha_7^2-\ha_6^2 > 0 $. Notice that in order to correspond to the
QCD partition function, the limit $\ha_k \to 0$ should be
regular.

We will now explicitly work out the generating function (\ref{ZSUSY})  
in the quenched case
where $N_f=0$. In order to do so we will use the parametrization Eq. (\ref{U})
which has a flat measure \cite{DOTV}. 
In this representation the generating function with $\hm>0$ 
and ${\rm Im }(\hz') < 0$ becomes
(after $\alpha$ and $\beta$ are integrated out)
\be
\label{ZSUSY2} 
 && Z^\nu_{1|1}(\hm,\hm',\hz,\hz';\ha_i) \\
& = & \int_{-\infty}^\infty ds\int_{-\pi}^\pi
\frac{d\theta}{2\pi} \ e^{(i\theta-s)\nu}
\exp[-\hm\sin(\theta)- i \hm'\sinh(s)+i\hz\cos(\theta)
  -i\hz'\cosh(s) \nn\\
&&
+4\ha_6^2(-i\sin(\theta)+\sinh(s))^2
+4\ha_7^2(\cos(\theta)-\cosh(s))^2
+2\ha_8^2(\cos(2\theta)-\cosh(2s))] 
 \nn\\
&& \times
\big(-\frac{\hm}{2}\sin(\theta)+i\frac{\hm'}{2}\sinh(s)+
i\frac{\hz}{2}\cos(\theta)+i\frac{\hz'}{2}\cosh(s)
\nn\\
&&
\quad -4(\ha_6^2+\ha_7^2)(\sin^2(\theta)+\sinh^2(s)) 
+2\ha_8^2(\cos(2\theta)+\cosh(2s)+e^{i\theta+s}+e^{-i\theta-s})\big).
\nn
\ee
The $s$-integral is convergent for $\ha_8^2-\ha_6^2-\ha_7^2>0$. In
this case it is easily checked numerically that
$Z_{1|1}(\hm,\hm,\hz,\hz;\ha)=1$, as required by the definition 
of the generating function. We discuss the convergence criterion in 
 detail in section \ref{subsec:conv} below.

\subsection{The Microscopic distribution of the chirality over the
  real eigenvalues of $D_W$}
\label{sec:real}

In this section we show that the parameter $\nu$
in the chiral Lagrangian is the index of the Dirac operator $D_W$.
Since we already have computed the generating function (\ref{ZSUSY})
the condensate defined in Eq. (\ref{Sigmadef}) can be expressed as
\be\label{SigmaQ}
\Sigma^\nu(\hm;\ha_i) & \equiv & 
-\lim_{{\hm}'\to \hm} \frac{d}{d\hm'}
Z^\nu_{1|1}(\hm,{\hm}',\hz=0,\hz'=-i\epsilon;\ha_i).  
\ee
This corresponds to a resolvent regulated with $i\epsilon \gamma_5$ instead
of $i\epsilon$, and as was discussed in section II, it has 
quite different properties with a discontinuity that  does not give the spectral
density but rather  the distribution  $\rho^\nu_\chi$ of the chirality 
over the real eigenvalues
of $D_W$ given in Eq. (\ref{distOFchi}).
It follows from (\ref{ZSUSY2}) that for Im$(\hz'=\hz)<0$
\be\label{SigmaQ2}
\Sigma^\nu(\hm;\ha_i) 
& = & -\int_{-\infty}^\infty ds\int_{-\pi}^\pi \frac{d\theta}{2\pi} \ 
\sin(\theta)e^{(i\theta-s)\nu} 
\exp[-\hm\sin(\theta)- i \hm\sinh(s)-\epsilon \cosh s \nn \\
&&
+4\ha_6^2(-i\sin(\theta)+\sinh(s))^2
+4\ha_7^2(\cos(\theta)-\cosh(s))^2
+2\ha_8^2(\cos(2\theta)-\cosh(2s))] \nn \\
&& \hspace{-1.5cm}\times 
\Big(-\frac{\hm}{2}\sin(\theta)+i\frac{{\hm}}{2}\sinh(s)
-4(\ha_6^2+\ha_7^2)(\sin^2(\theta)+\sinh^2(s))
\nn\\
&& 
+2{\ha}_8^2(\cos(2\theta)+\cosh(2s)+e^{i\theta+s}+e^{-i\theta-s})+\frac{1}{2}\Big).
\ee
The discontinuity of this resolvent is equal to microscopic limit
of the distribution of the chirality over the real eigenvalues of $D_W$ 
defined in Eq. (\ref{distOFchi}), 
\be
\label{rhoWreal}
\rho_\chi^\nu(\lambda^W;\ha_i) =\left\langle\sum_{\lambda_k^W \in
 {\mathbb R}} \delta(\lambda_k^W+ \lambda^W) \, {\rm sign}[\langle k|\gamma_5|k\rangle]\right\rangle
 = \frac{1}{\pi} {\rm Im}[\Sigma^\nu(\hm=\lambda^W;\ha_i)].
\ee
It can be verified numerically that with the resolvent (\ref{SigmaQ2})
\be
\int \rho_\chi^\nu(\lambda^W;\ha_i) d\lambda^W = \nu.
\ee
This demonstrates that the sectors 
introduced in Eq.~(\ref{Zfull}) correspond to a Dirac operator $D_W$
with index $\nu$, as defined in (\ref{defIndex}).

\begin{center}
\begin{figure}
\includegraphics[width=8.5cm,angle=0]{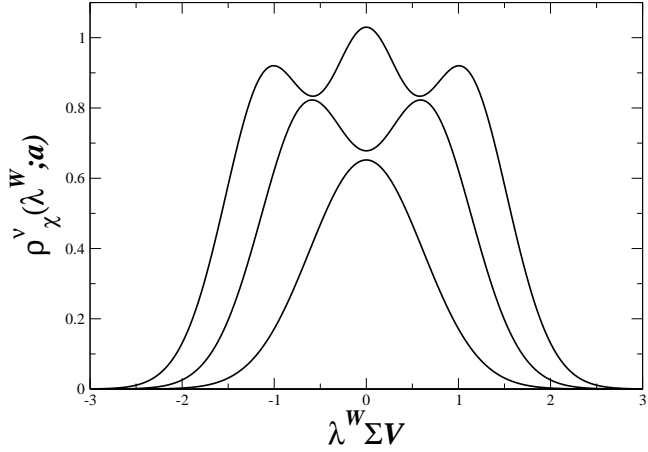}
\caption{\label{fig:rhotopo} The quenched distribution of the
  chirality over the real modes of $D_W$ 
  plotted for $\nu=1,2$ and $3$ with $\ha_8=0.2$. For the plot we have
  chosen $\ha_6=\ha_7=0$. The real modes
  repel each other, making the preferred locations of the  
  eigenvalues clearly visible.}
\end{figure}
\end{center}

In Fig.~\ref{fig:rhotopo} we plot the distribution of the chiralities
over the real eigenvalues of $D_W$, $\rho_\chi^\nu(\lambda^W, \ha_i)$,
for $\nu=1,2,3$ for $\ha^2_8 =0.2$ and $\ha_6= \ha_7 = 0$. For  
this value of $\ha_8$ additional pairs of real eigenvalues appear only
rarely and $\rho_\chi^\nu(\lambda^W, \ha_i)$ is a good approximation
to the density of the real modes of $D_W$.  
Indeed, the presence of $\nu$ real eigenvalues is
 clearly visible. As we will show below, we can
make an even more precise description of this in the limit of very
small $\hat{a}_8$. In that limit we have $\nu$ real modes which 
turn out to be described
by the $|\nu| \times |\nu|$ Gaussian Unitary Ensemble of Random
Matrix Theory.

\subsection{The Microscopic Spectrum of $D_5$}

In the continuum ($a=0$) the eigenvalues $i\lambda_k$ of $D_W$ and
eigenvalues $x_k$ of $D_5$ are related to each other
through $(\lambda_k^W)^2=(\lambda_k^5)^2-m^2$. In this case, the
microscopic eigenvalue density of $D_5$,  
\be\label{rho5a0}
\rho_5^\nu(\lambda^5>\hm,\hm;\ha_i=0) 
= \frac{\lambda^5}{\sqrt{(\lambda^5)^2-\hm^2}} 
\rho^\nu(\sqrt{(\lambda^5)^2-\hm^2})
\ee 
follows from that of $D_W$. In the quenched case we have \cite{RMT,DOTV}
\be
\rho^\nu(\lambda) = \frac{\lambda}{2}
\left[J_\nu(\lambda)^2-J_{\nu+1}(\lambda)J_{\nu-1}(\lambda)\right].
\ee

\begin{center}
\begin{figure}
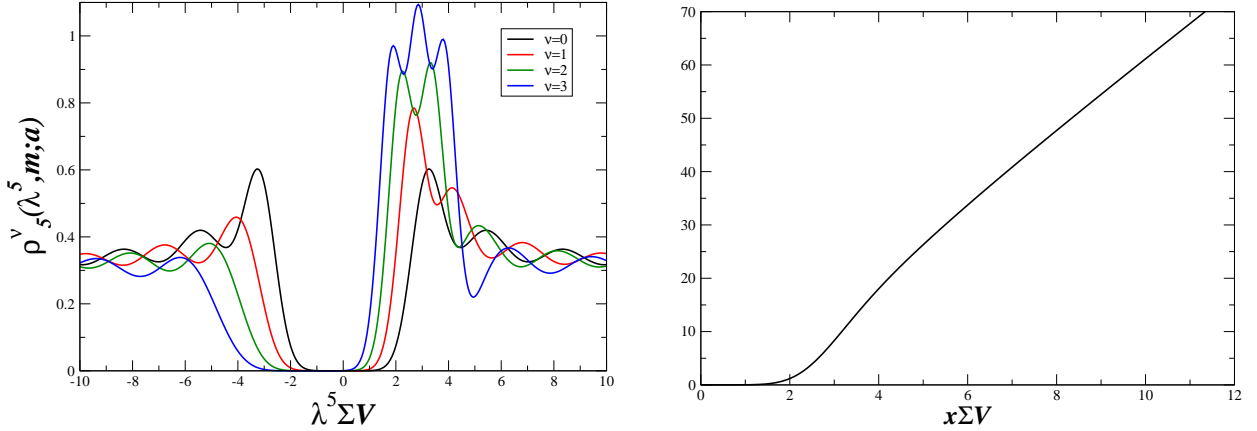

\includegraphics[width=8.1cm,angle=0]{rho5nu01234a0p2m3.eps}\hfill
\includegraphics[width=7.5cm,angle=0]{rho5sumnuint.eps}
\caption{\label{fig:rho5} {\bf Left:} The microscopic spectral density of $D_5$
  for $\nu=0,1,2$ and $3$ with, $\hm=3$, $\ha_8=0.2$ and $\ha_6=\ha_7=0$. 
    At nonzero value of the
  lattice spacing the zero modes spread out into a region around
  $\hx=\hm$. For negative values of $\nu$ the spectral density is reflected
  at the origin. {\bf  Right:} After summation over $\nu$ with a
  Gaussian weight and integrating from $\lambda^5=0$ up to $\lambda^5=x$. 
  That is, the average number of eigenvalues below $x$. For the plot
  we have chosen $\langle\nu^2\rangle=1$. The values of $\hm$ and
  $\ha_i$ are as in the left figure. Compare with Fig.~1 of \cite{LP}.}
\end{figure}
\end{center}

To obtain the eigenvalue density, $\rho_5^\nu$, of the
Hermitian Wilson Dirac operator $D_5$ for non-zero values of $\ha$ we
first evaluate the resolvent 
\be 
\label{G}
G^\nu(\hz,\hm;\ha_i) & \equiv & 
-\lim_{{\hz}'\to \hz} \frac{d}{d\hz'} Z^\nu_{1|1}(\hm,\hm,\hz,{\hz}'-i\epsilon;\ha_i)
\ee 
to find 
\be
\label{res-result}
G^\nu(\hz,\hm;\ha_i) & = & \int_{-\infty}^\infty ds\int_{-\pi}^\pi
\frac{d\theta}{2\pi} \
i\cos(\theta)e^{(i\theta-s)\nu} \\
&& \hspace{-1cm}\times\exp[-\hm\sin(\theta)- i \hm\sinh(s)+i\hz\cos(\theta)
  -i(\hz-i\epsilon) \cosh(s) \nn \\
&&\hspace{-1cm}
+4\ha_6^2(-i\sin(\theta)+\sinh(s))^2
+4\ha_7^2(\cos(\theta)-\cosh(s))^2
+2\ha_8^2(\cos(2\theta)-\cosh(2s))] 
 \nn \\
&& \hspace{-1cm}\hspace{-2cm}\times
\big(-\frac{\hm}{2}\sin(\theta)+
i\frac{\hm}{2}\sinh(s)+i\frac{\hz}{2}\cos(\theta)+i\frac{\hz}{2}\cosh(s)
\nn\\ 
&&\hspace{-1cm}
-4(\ha_6^2+\ha_7^2)(\sin^2(\theta)+\sinh^2(s))
+2\ha_8^2(\cos(2\theta)+\cosh(2s)
+e^{i\theta+s}+e^{-i\theta-s})+\frac{1}{2}\big) . \nn
\ee
The microscopic spectral density of $D_5=\gamma_5(D_W+m)$ in the quenched limit
can then be expressed in terms of the imaginary part
\be
\label{rho5}
\rho^\nu_5(\lambda^5,\hm;\ha_i) & = & \frac{1}{\pi}{\rm Im} \ G^\nu(-\lambda^5,\hm;\ha_i).
\ee
In Fig. \ref{fig:rho5} this density is plotted for four values of
$\nu$ for fixed $\ha$ and $\hm$. The similarity of
$\rho^\nu_\chi(\lambda^W)$ and $\rho^\nu_5(\lambda^5)$ for
$\lambda^W\sim\lambda^5-\hm$ (see Figure \ref{fig:topo5} for a direct
comparison) is not accidental: As we show in section 
\ref{sec:small_a}, for $\ha_8\ll1$ the $\nu$ real modes of $D_W$ are mapped directly 
to $\nu$ modes in the vicinity of $\hm$, see 
Eq.~(\ref{rho5rhoWreal-map}).  

It is of course also possible to sum over all sectors with a given
index, for an example see the right hand panel of \ref{fig:rho5}.

\begin{center}
\begin{figure}[t]
\includegraphics[width=7.5cm,angle=0]{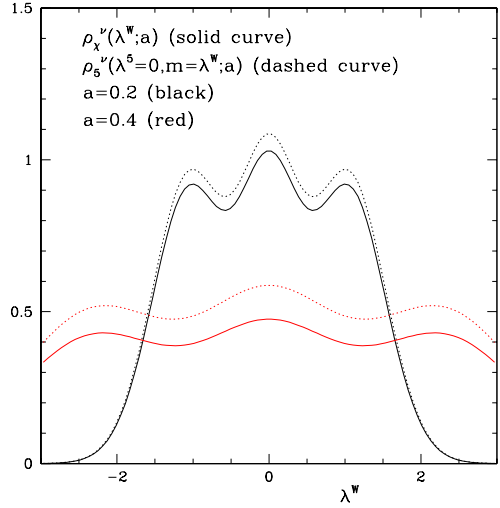}
\caption{\label{fig:dwreal} 
The distribution of the chirality over the real eigenvalues of $D_W$
$\rho_\chi^{\nu=3}(\lambda^W;\ha_i)$ 
(solid curves) and $ \rho_5^{\nu=3}(\lambda^5=0,\hm=\lambda^W;\ha_i)$
as a function of $\lambda^W$.  
The values of $\ha_8 $ (here denoted by $a$) are given in the legend
of the Figure ($\ha_6=\ha_7=0$ here). The density of the real
eigenvalues of $D_W$ 
is bounded between the solid and the dashed curve.}
\end{figure}
\end{center}

In Section II we discussed the mass dependence of 
$\rho_5^\nu(\lambda^5 = 0, \hm; \ha_i)$. The  density of the real
eigenvalues of $D_W$ satisfies the inequality (\ref{rhoINEQ}).
In Fig. \ref{fig:dwreal}  we plot both sides of this inequality as a
function of $\lambda^W$.

\subsection{The effect of $W_6$ and $W_7$}
\label{subsec:W6W7inWCPT}

In the derivation above we have explicitly included the effect of
$W_6$ and $W_7$ in the Lagrangian and then performed the relevant
super-traces and fermionic integrations. Here we point out an
alternative way to compute the effect of $W_6$ and $W_7$. Not only
will this give us a direct way to quantify the effect of $W_6$ and
$W_7$ on the spectrum of the Dirac operator, it also provides a 
simple way
to include the effects of these terms in the chiral
Random Matrix Theory discussed below.

Since $W_6$ and $W_7$ are coupling constants of double-trace terms,
we can re-express the microscopic partition function (\ref{Zfull}) as
\be\label{partfW6W7}
Z^\nu_{N_f}(\hm,\hz;\ha_6,\ha_7,\ha_8) 
= \frac{1}{16 \pi \ha_6 \ha_7}\int_{-\infty}^\infty dy_6 dy_7 \ 
e^{-\frac{y_6^2}{16|\ha_6^2|}-\frac{y_7^2}{16|\ha_7^2|}} \ 
Z_{N_f}^\nu(\hm-y_6,\hz-y_7;0,0,\ha_8) .  
\ee
Here we have written the expression valid for negative values $W_6$
and $W_7$. If $W_6>0$ the shift of $\hm$ is instead
along the imaginary axis, $\hm-iy_6$. For $W_7>0$ we analogously
shift $\hz$ along the imaginary axis, $\hz-iy_7$.

Since exactly the same rewriting is valid for the generating function 
(\ref{ZSUSY}) also the quenched spectral density at non-zero values of 
$W_6$ and $W_7$ follows from that with $W_8$ alone
\be
\label{intW6W7}
\rho^\nu_5(\lambda^5,\hm;\ha_6,\ha_7,\ha_8) = 
\frac{1}{16 \pi \ha_6 \ha_7}\int_{-\infty}^\infty dy_6 dy_7 \ 
e^{-\frac{y_6^2}{16|\ha_6^2|}-\frac{y_7^2}{16|\ha_7^2|}} \ 
\rho^\nu_5(\lambda^5-y_7,\hm-y_6;0,0,\ha_8).
\ee
Interestingly,
for $W_6<0$ the quenched spectral density of $D_5$ is a Gaussian average
of spectra with a smooth distribution of quark masses. 
The integrals over $y_6$ and $y_7$ in (\ref{partfW6W7}) and
  (\ref{intW6W7}) can only be interchanged with the noncompact
  integral in the partition function if the inequality
$\ha_8^2-\ha_6^2 -\ha_7^2 > 0 $ is satisfied. Furthermore, the shift 
must be along the real axis in order that the discontinuity across the
real axis remains linked to the eigenvalue density.

\section{Limiting Cases}
\label{sec:small_a}

In this section we discuss various limiting cases of the spectral
density of $D_5$ and $D_W$. 
Since the Dirac spectrum
at $W_{6, 7} \ne 0 $ is given by a Gaussian integral over the
Dirac spectrum for $W_6=W_7 =0$, only the dependence on $W_8$ will be
analyzed in this section.

\subsection{Small $\ha_8$-limit for $\hz-\hm$ fixed}

We first show explicitly that in the limit $\ha_8\to 0$ the microscopic resolvent
(\ref{res-result}) reduces to the known analytical result of the
continuum limit. 
To derive the $\ha_8 \to 0$ limit of the microscopic spectral density, we
use the integrals
\be
\int_{-\pi}^\pi \frac{d\theta}{2\pi}  e^{i\nu \theta} e^{-\hm\sin\theta +i\hz\cos\theta} &=& \left ( \frac {\hz-\hm}{\hz+\hm} \right )^{\nu/2} 
I_\nu(i\,{\rm sign}(z) \sqrt{\hz^2-\hm^2}),
\label{iint}
\ee
\be
&& \hspace{-30mm}\lim_{\ha_8\to 0}\int_{-\infty}^\infty ds e^{-\nu
  s}e^{-i\hm\sinh(s)-i(\hz-i\epsilon)\cosh(s)-2\ha_8^2\cosh(2s)} \\
&=&2 \left ( \frac {\hz-i\epsilon-\hm}{\hz-i\epsilon+\hm} \right )^{-\nu/2} K_\nu(i \sqrt{(\hz-i\epsilon)^2-\hm^2}),\nn
\label{kint}
\ee
valid at fixed  $|\hm-\hz| \ne 0$, and obtain
\be
\lim_{\ha_8\to0}G^\nu(\hz,\hm;\ha_8) &=& -\frac 14 K_\nu(\eta)[ (\hz+\hm)(\beta^2 I_{\nu+2}(\eta) + I_\nu(\eta))
+(\hz-\hm)(\beta^{-2} I_{\nu-2}(\eta) + I_\nu(\eta))]\nn\\
&&-\frac 14 [\beta I_{\nu+1}(\eta) +\beta^{-1} I_{\nu-1}(\eta)][(\hz+\hm)\beta K_{\nu-1}(\eta)
+(\hz-\hm) \beta^{-1} K_{\nu+1}(\eta)] \nn \\
&& +\frac i2 K_\nu(\eta)[\beta I_{\nu+1}(\eta) +\beta^{-1} I_{\nu-1}(\eta)],
\ee
with 
\be
\eta&=& i \sqrt{\hz^2-\hm^2},\qquad
\beta= \sqrt{\frac{\hz-\hm}{\hz+\hm}}.
\ee
We now use the recursion relations
\be
I_{\nu+2}(\eta) &=& I_\nu(\eta) -\frac {2(\nu+1)}\eta I_{\nu+1}(\eta),\nn \\
I_{\nu-2}(\eta) &=& I_\nu(\eta) +\frac {2(\nu-1)}\eta I_{\nu-1}(\eta),\nn \\
K_{\nu+1}(\eta) &=& K_{\nu-1}(\eta) +\frac{2\nu}\eta K_\nu(\eta),
\ee
and
the Wronskian identity to obtain the resolvent of $D_5$ in the limit
$\ha_8\to0$ for $\hz-\hm$ fixed
\be
\label{G_a-small}
G^\nu(\hz,\hm) =-z(I_\nu(\eta) K_\nu(\eta) + I_{\nu+1}(\eta)K_{\nu-1}(\eta)) + \frac \nu{\hz-\hm}.
\ee
As we will now demonstrate, this is the resolvent of $D_5$ for $a = 0$.
The resolvent of $D_W$ at  $a=0$ can be expressed as
\be
\left\langle {\rm Tr} \frac 1{D_W+z}\right\rangle = \left\langle \sum_{\lambda^W_k >0}
\frac {2z}{(\lambda_k^W)^2 +z^2}\right\rangle + \frac \nu z.
\ee
The eigenvalues of $D_5$ are $\lambda_k^5 = \pm \sqrt{(\lambda_k^W)^2 +m^2}$
for $\ha_8=0$ resulting in the resolvent
\be
 &&  \left\langle{\rm Tr}  \frac 1{D_5+z} \right\rangle
= \left\langle  \sum_{\lambda^W_k >0}\frac {2z}{z^2-(\lambda_k^W)^2 -m^2} \right\rangle+ \frac \nu {z-m} \nn \\
&=& -\left\langle  \sum_{\lambda^W_k >0}\frac {2z}{(\lambda^W_k)^2  +m^2-z^2} \right\rangle
+ \frac \nu {z-m} \nn \\
&=& -z \left . \frac{\Sigma^\nu_{\lambda^W_k> 0}(x)}{x}\right |_{x =i\sqrt{z^2 -m^2}} +
 \frac \nu {z-m}.
\ee 
Inserting the $\ha_8=0$ result for $\Sigma^\nu$, see eg.~Eq.~(45) of 
\cite{DOTV}, we consistently find (\ref{G_a-small}) above.
We conclude that in the limit $\ha_8\to 0$ the eigenvalue density
$\rho_5^\nu(\lambda^5,\hm;\ha_8)$ is given by the standard $\ha_8=0$
result (\ref{rho5a0}) as long as $|(\lambda^5 -\hat{m})|/\hat a_8  \gg 1$.

\subsection{Small $\ha_8$-limit for $(\hz-\hm)/\ha_8$ fixed}

As we have just seen, for $a = 0$ the contribution of the zero modes
of $D_W$ to the resolvent of $D_5$ is given by
\be
\frac \nu {\hz-\hm}.
\ee
For $ a\ne 0$ the zero modes are smeared out so that the singularity in
the resolvent is smoothened, and we expect the imaginary part of the resolvent
to behave as
\be
{\rm Im} [G^\nu (\hz,\hm)] \sim\frac 1\ha_8 F(\hz-\hm),
\ee
with $F$ a peaked function. 
In \ref{app:A} we extract the analytical expression for $F$ from the
small $\ha_8$ limit of the microscopic result for the resolvent given in
Eq. (\ref{res-result}). Amazingly, the result is given by the eigenvalue
density of a $|\nu|\times|\nu|$ Hermitian Random Matrix Theory
\be
\rho_5^\nu(\lambda^5,\hm;\hat{a}_8\ll1) & = &
\frac{1}{4\ha_8}\exp(-u^2)\sum_{k=0}^{\nu-1} \frac{H_k(u)H_k(u)}{2^k k!
  \sqrt{\pi}}  \\
& = & \frac{1}{4\ha_8}\frac{e^{-u^2} }{2^{\nu}(\nu-1)!\sqrt{\pi}}
[H_\nu^2(u) -H_{\nu+1}(u)H_{\nu-1}(u)], \nn
\ee 
with
\be
u = \frac{\lambda^5-\hat{m}}{4\hat{a}_8}.
\ee 
This is the familiar spectral density of the $\nu\times\nu$ Gaussian Unitary
Ensemble shifted by $\hm$ and rescaled by $1/4\ha_8$. For $\nu = 1$ the
result is a simple Gaussian.

For the normalization of the Hermite polynomials we used the convention
\be
\int_{-\infty}^\infty dx \ H_j(x)H_k(x) e^{-x^2} =\delta_{jk}2^k k! \sqrt{\pi}.
\label{normH}
\ee 
The $1/\ha_8$ contribution to the spectral density is therefore normalized
to $\nu$. For $\nu =0 $ the leading small-$\ha_8$ corrections are
$O(\log(\ha_8))$, see \ref{app:A} for details, and will not be considered
to the order we are working.

\subsection{For small $\ha_8$ the distribution $\rho_\chi^\nu$ is given by $\rho^\nu_5$}

The distribution of the chirality over the real eigenvalues of $D_W$,
$\rho_\chi^\nu(\lambda^W;\ha_8)$, is obtained from the resolvent of
$D_5$, Eq. (\ref{res-result}), by replacing the pre-exponential factor
$ i\cos \theta \to -\sin\theta$ and putting $z \to 0$. 
As discussed in section \ref{sec:real}, to leading order in $\ha_8$, 
this is equal the spectral density of the
real eigenvalues of $D_W$. For small $\ha_8$ the width of
the spectrum $\sim \ha_8$ and we can thus consider the limit of small $\ha_8 $
with $ \hm/\ha_8$ fixed. This is exactly the limit that was considered
in previous section.

\begin{center}
\begin{figure}
\includegraphics[width=8.5cm,angle=0]{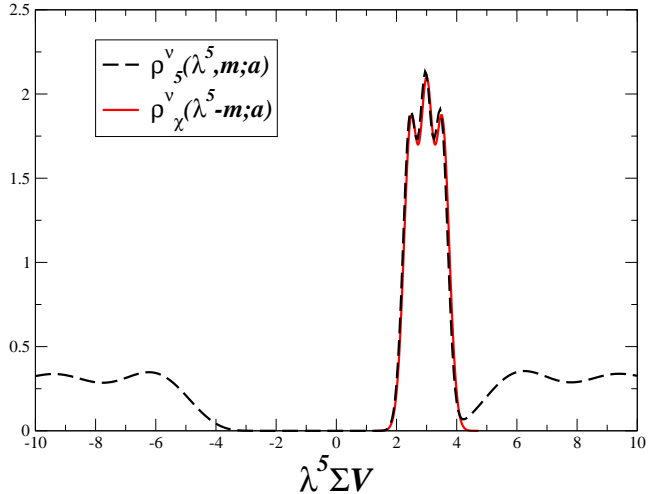}
\caption{\label{fig:topo5} The microscopic spectral density of $D_5$
  for $\nu=3$ with $\hm=3$ and $\ha_8=0.1$  and $\ha_6=\ha_7=0$. Also plotted
  is the distribution of the chirality over the real modes of $D_W$
  shifted by the mass, that is 
  $\rho^\nu_\chi(\lambda^5-\hm;\ha_8)$. Since the real modes are almost
  chiral for small $\ha_8$ the two distributions merge.}
\end{figure}
\end{center}

 To leading order, only the negative exponent of
$-\sin \theta$ has to be taken into account (which, up to a minus sign,
is the same as the negative exponent of  $i\cos \theta$). We
thus find 
\be\label{rho5rhoWreal-map}
 \rho_{\rm real}^\nu(\lambda^5-\hm; \ha_8)  
= \rho_5^\nu(\lambda^5,\hm;\ha_8)\qquad  {\rm for}
 \quad \ha_8 \to 0 \quad {\rm and } \quad (\hm-\lambda^5)/\ha_8 \quad {\rm fixed}.  
\ee
This fact is also demonstrated graphically in Figure
\ref{fig:topo5}. The two distributions merge because the real modes
are almost chiral. To see this we start with a  real mode of $D_W$
\cite{Itoh}   
\be
D_W \phi_j = \lambda^W_j \phi_j.
\ee
It follows that
\be
D_5\phi_j=\gamma_5(D_W +m) \phi_j = (\lambda^W_j+m) \gamma_5 \phi_j.
\ee
Now if the real modes of $D_W$ are chiral $\gamma_5 \phi_j =\pm\phi_j$
then the $\nu$ real eigenvalues of $D_W$ are mapped onto $\nu$ 
eigenvalues of $D_5$ with a trivial shift by $m$. More precisely,
\be
\gamma_5\phi_j = 
\langle j |\gamma_5 | j\rangle\phi_j + \delta \phi \qquad {\rm with} 
 \quad \langle j| \delta \phi \rangle =0 \quad {\rm and}
\quad \delta \phi \sim O(a), 
\ee
so that
\be
D_5\phi_j = (m+\lambda^W_j) 
\langle j |\gamma_5 | j\rangle\phi_j +O(a).
\ee
Since the distributions of the two merge in 
the small $\ha_8$ limit, (see Eq.~(\ref{rho5rhoWreal-map})), this
explicitly confirms that the chirality of the real modes is unity to
leading order in $\ha_8$.

\subsection{Scaling of Smallest Eigenvalue}

For $\nu=1$ and $\ha_8\ll1$ the distribution of the single real
eigenvalue of $D_W$ takes the Gaussian form
\be
\rho^{\nu=1}_{\rm real}(\lambda^W;\ha_8\ll1) = \frac 1{4\ha_8\sqrt \pi}
\exp\left(-\frac {(\lambda^W)^2}{16\ha_8^2}\right). 
\ee
In physical units the width of the distribution is
\be
\sigma = \frac{ \sqrt{ 8 a^2 W_8}}{\Sigma \sqrt V}.
\ee
This is also the width parameter of the Gaussian tail of the spectral
density of $D_5$ inside the gap. 
In lattice simulations of the Wilson Dirac operator, a 
scaling of the width of the
distributions of the smallest eigenvalue $\sim a /\sqrt V$ has been 
reported \cite{Luscher} for $N_f=2$ simulations.

It is also instructive to consider the ratio of the width parameter and
the average spacing $\Delta\lambda=\pi/\Sigma V$ of the Dirac
eigenvalues at $a=0$,
\be
\frac \sigma{\Delta \lambda} = (8a^2 W_8 V)^{1/2} \frac{1}{\pi}
= \ha_8 \frac{\sqrt{8}}{\pi}.
\ee
This gives an intuitive interpretation of the dimensionless low-energy
constant $\ha_8$.

\subsection{ Mean Field Limit}
\label{mft}
For large  $\hm$, $\hz$, and $\ha_i^2 $ the graded generating function can 
be evaluated by a saddle point approximation. This limit
corresponds to the lowest non-trivial
order in the usual perturbative expansion ($p$-regime) as considered
in \cite{Sharpe2006}. 
We will focus, in particular, on the behavior of the spectral density 
 $\rho_5$ near the edge of the spectrum.
At mean field level the dependence on the index of the Dirac operator 
is suppressed and we will therefore start from the $\nu=0$ expression
and drop the index $\nu$ below. Since 
the Dirac spectrum
at $W_{6, 7} \ne 0 $ is given by a Gaussian integral over the
Dirac spectrum for $W_6=W_7 =0$, only the dependence on $W_8$ will be
taken into account in this section.

The expression for the spectral density can be written
as
\be
\rho_5^{\nu=0}(\hm,\hz;\ha_8) = \frac 1\pi {\rm Im} 
\int_{-\infty}^\infty ds \int_{-\pi}^\pi\frac{d \theta}{ 2 \pi} 
e^{S_f(\hm,\hz;\ha_8)} \frac d{dz'} e^{S_b(\hm,\hz'; \ha_8)}
P(\hm,\hz,\hz';\ha_8) \big|_{\hz'=\hz},
\ee
where $S_f$, $S_b$ and $P$ can be read off from Eqs. (\ref{G}),
(\ref{res-result}). By shifting integration contours  according to
\be
\theta &=& i(-r+\frac {i\pi}2), \nn \\
s &=& t- \frac{\pi i}2,
\label{newvars}
\ee
the fermionic and bosonic exponents become real and equal up to a sign
\be
\tilde S_f(r) &=& \hm\cosh r + \hz \sinh r - 2\ha_8^2 \cosh 2r,\nn\\
\tilde S_b(t) &=& -\hm \cosh t - \hz \sinh t + 2 \ha_8^2 \cosh 2t.
\label{saddles} 
\ee
Also the prefactor in terms of these variables
\be
P(s,r) =-\frac i2 [\hm \cosh r + \hz \sinh r + \hm \cosh t +\hz \sinh t  -2
 \ha_8^2(( e^r +e^t)^2 
+(e^{-r} + e^{-t})^2)]
\label{pref}
\ee
becomes real (an overall factor of $i$ is included in the
fermionic integration).

For large $\hm $ and $\hz$ the integrals can be evaluated by a saddle
point approximation. 
  It is convenient to introduce $u=\sinh s$ as new variable so that 
the potential
(\ref{saddles}) is given by
 \be
S_b(u) = -\hm \sqrt{1+u^2} - \hz u +2\ha_8^2 (2u^2+1). \label{saddleu}
\ee
The edge of the spectrum is the point where two real saddle points
coalesce and move into the complex plane. At this point
\be
S'_b(u_g) = 0, \qquad {\rm and} \qquad  S_b''(u_g) = 0,
\ee
where $ S_b(u)$ ia given in Eq. (\ref{saddleu}).
The second equation results into
\be
u_g^2 = \left ( \frac \hm{8\ha_8^2}\right )^{2/3}-1,
\ee
which combined with the vanishing first derivative leads to the mean
field result for the gap 
\be
\hz_g = \hm \left ( 1 - \left ( \frac {8\ha_8^2}\hm\right
)^{2/3}\right )^{3/2}. 
\ee
The mean field 
spectrum is symmetric around the origin and above we displayed the
positive solution for the gap.

Depending on the position of $\lambda^5$ with respect to the position of
the gap $\hz_g$
we can distinguish three parameter domains:

\begin{itemize}

\item[i)] $|\lambda^5| < z_g $, then all saddle points in terms of the
$r$ and $t$ variables are real. The leading saddle point determines the
fermionic integral but does not contribute to the imaginary part of the
bosonic integral (which gives the imaginary part of the resolvent).
The imaginary part is given by a subleading saddle point when combined
with the fermionic contribution gives an exponentially suppressed tail.
Combining this with the fermionic integral gives the spectral density.
In the small $\hat{a}_8$ limit, when $\cosh s \approx \sinh s $ at the saddle point,
the bosonic integral becomes Gaussian resulting in the Gaussian tail
\be
\rho_5(\lambda^5,\hm; \ha_8) \sim 
\exp\left(\hm-\frac{\hm^2}{2(\hm-8\ha_8^2)} -\frac{(\hm-\lambda^5)^2}{16\ha_8^2}\right).
\ee

\item[ii)] $|\lambda^5| > z_g $, then a pair of real saddle points has turned into a pair
of complex conjugate saddle points. Only one of the saddle points can be
accessed by the bosonic integration contour, but they both contribute to 
the fermionic integral. When the saddle points of the bosonic and fermionic integrals
are the same, the exponents
cancel resulting in a smooth contribution to the spectral density. When 
they are different, they result in an oscillatory exponent which is subleading
because of  the prefactor. It is instructive to work out the case
$\ha_8 = 0$ which is done in \ref{app:B}. Also the non-zero $\ha_8$ case is
discussed in this appendix.

\item[iii)] $\lambda^5\approx z_g$, then two saddle points are close and the exponents
can be approximated by a cubic polynomial.  This is the scaling domain where
$V(z-z_g)^{3/2}$  is kept fixed in the thermodynamic limit for fixed $m$ and $a$. 
In principle, the exact generating function can be evaluated in this limit, and
according to universality arguments it should give a spectral density that
can be expressed in terms of Airy functions as (see
Eq. (\ref{defDelta}) for the definition of $\Delta$) 
 \be
\rho_5(x) = \frac 1\Delta ({\rm Ai}'(x)^2 - x {\rm Ai}^2(x)), 
\qquad {\rm with} \qquad
x= (z_g-z)/\Delta. 
\label{airy}
\ee
The leading
order asymptotic behavior of the Airy function on both sides 
of $z_g$ follows from a saddle point approximation of the generating function
in this domain. This is shown in \ref{app:B}. 

\end{itemize}

\begin{center}
\begin{figure}
\includegraphics[width=8.5cm,angle=0]{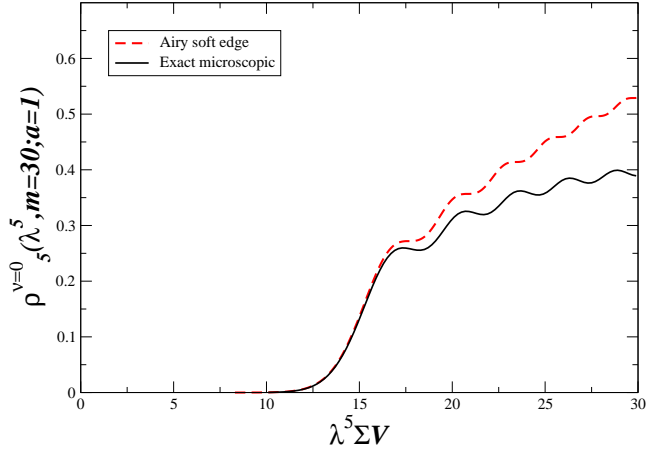}
\caption{\label{fig:Airy} The Airy behavior of $\rho_5^\nu(\lambda^5, \hz)$ 
    dots for    $\hm =30$, $\ha_8 =1$  and $\lambda^5$ close to $z_g$
    vs. Eq. (\ref{airy}).} 
\end{figure}
\end{center}

\subsection{Edge Scaling}

As discussed in previous section,
the supersymmetric generating function can be evaluated in a counting
scheme where $\hm\gg 1$, $\ha_8^2\gg 1$ and $(\hz-\hz_g)^{3/2}\gg 1$ 
are all  of the same order. This is a universal scaling domain where 
the average spectral density is given by the expression in terms of
Airy functions (see Eq. (\ref{airy})). We illustrate this in
Fig.~\ref{fig:Airy} 
for $\hm = 30$ and $\ha_8 =1$.

The  distribution, $p_{\rm min}$,
of the smallest positive eigenvalue 
for the corresponding density  Eq. (\ref{airy})
is also known. 
It is given by the Tracy-Widom distribution \cite{Tracy,For,NWch6}
for $\beta =2$,
\be
p_{\rm min}(z) = \frac d{dz} F_2((z-z_g)/\Delta),\\
F_2(x) =  e^{-\int_{-\infty}^x(x-y) q^2(y) dy},
\ee
where $q(x)$ is the solution of
\be
q''(x) = - x q(x) + 2 q^3(x)
\ee
with boundary condition that $q(x) \to {\rm Ai}(-x) $ for $x \to -\infty$.

\section{Random Matrix Theory for the Wilson Dirac Operator}
\label{sec:WRMT}

A chiral Random Matrix Theory for lattice QCD with Wilson
fermions is constructed from  the most
general $\gamma_5$-Hermitian matrix. This random matrix Wilson Dirac
operator has the block structure 
\be
\tilde{D}_W= \mat \tilde a A & iW \\ iW^\dagger & \tilde a B \emat,
\label{w-diracOP}
\ee
where  
\be
A=A^\dagger \quad {\rm and}  \quad B^\dagger = B
\ee
are $(n+\nu) \times (n+\nu)$ and $n \times n$ complex matrices, 
respectively, and $W$ is an arbitrary complex $(n+\nu)\times n$ matrix. In
Eq.~(\ref{w-diracOP}) and below we 
use tildes to indicate quantities in the Random Matrix Theory which
are analogues of those in the quantum field theory. 
We relate the two sets in Eq.~(\ref{match}). 

We take the matrix elements to
be distributed with Gaussian weight
\be
\label{P}
{\cal P}(A,B,W) \equiv e^{-\frac {N}{4}{\rm Tr}[A^2+B^2] 
-\frac N2 {\rm Tr} [ W W^\dagger]},
\ee
where $N=2n+\nu$. Because of universality, results in the microscopic domain 
should not depend on the details of the probability distribution
\cite{RMT2}, but for simplicity we  take a Gaussian distribution.
The partition function of the Wilson chiral
Random Matrix Theory is then defined as
\be
\label{ZWRMT}
\tilde{Z}^\nu_{N_f} = \int dAdBdW \ \prod_{f=1}^{N_f}
\det(\tilde{D}_W+\tm_f+\tz_f\tilde{\gamma}_5) \ {\cal P}(A,B,W). 
\ee
The matrix integrals are over the complex Haar measure and 
$\tilde{\gamma}_5 = {\rm diag}(1,\cdots, 1,-1, \cdots, -1)$ is a
diagonal matrix with $n+\nu$ diagonal matrix elements equal to 1 and
$n$ diagonal matrix elements equal to $-1$. 

For large quark mass the matrix $\tilde{D}_5\equiv\gamma_5 (\tilde
D_W+m)$ has $n+\nu$ positive eigenvalues and $n$ negative eigenvalues, whereas
for large negative mass $\tilde{D}_5$ has $n+\nu$ negative eigenvalues and
$n$ positive eigenvalues.
Therefore at least $|\nu|$ spectral flow curves of $\tilde{D}_5$ have to cross
zero at least once. Since each crossing corresponds to a real
eigenvalue of $\tilde{D}_W$ we conclude that $\tilde D_W$ has at least $|\nu|$ 
real eigenvalues. The block structure of the matrix (\ref{w-diracOP})
guarantees that the Random Matrix Dirac operator has index $\nu$.

A chiral Random Matrix Theory for {\em staggered} fermions at 
finite lattice spacings was introduced in \cite{James}. 

\begin{figure}[t!]
  \unitlength1.0cm
    \epsfig{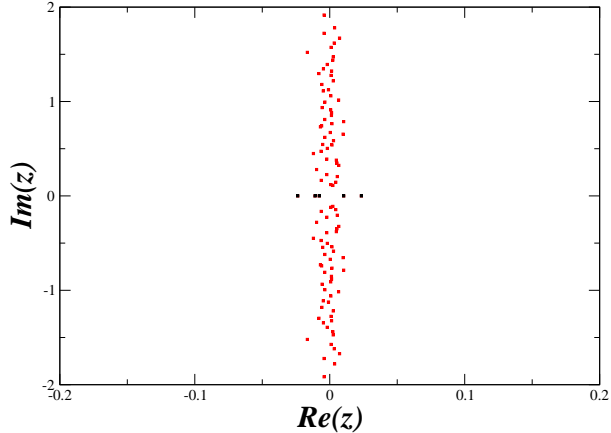} 
  \caption{\label{fig:WRMTscat} Scatter plot of the eigenvalues of
  $\tilde{D}_W$ with $\nu=5$ as obtained from Wilson chiral
  Random Matrix Theory.  The  real eigenmodes (black dots) are clearly
  visible. The width of the strip of complex eigenvalues in the 
  thermodynamic limit equals $|{\rm Re}(z)|<8a^2W_8/\Sigma$, as also
  follows from the chiral Lagrangian. See Eq. (\ref{match}) for the match
  between the parametrs of Wilson chiral Random Matrix Theory and
  those of the chiral Lagrangian.} 
\end{figure}

\subsection{From Wilson chiral Random Matrix Theory to Wilson chiral Perturbation Theory}
\label{subsec:WRMTtoWCPT}

\begin{figure}[t!]
  \unitlength1.0cm
  \epsfig{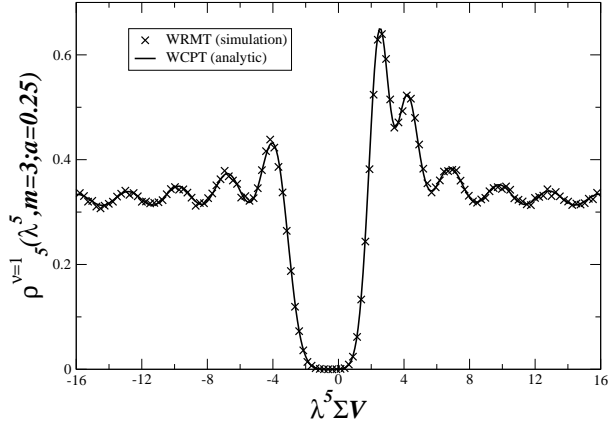}
  \caption{\label{fig:WRMTnumcheck}  The average
  spectral density of the hermitian Random Matrix Dirac operator 
$\tilde{D}_5$ with 
  $n=50$ and $\nu=1$ over an ensemble of 100,000 matrices (marked by
  crosses) and the prediction from Wilson Chiral Perturbation Theory.} 
\end{figure}

We now consider the microscopic limit of the chiral Random Matrix 
Theory for Wilson fermions in which
\be
\label{scaling}
\tm \sim N^{-1}, \qquad \tz\sim N^{-1}, \qquad \ta
\sim N^{-1/2}. 
\ee
We have studied numerically the quenched eigenvalue spectrum of the 
random matrix
Wilson Dirac operator, $\tilde{D}_W$, in this scaling regime. 
We find that the number of real eigenvalues of $\tilde{D}_W$ is equal 
to $\nu + O(\tilde{a}^4)$ for $\nu \ne 0$ and $\nu+O(\tilde{a}^2)$ for
$\nu = 0$, while the  
remaining eigenvalues come in complex conjugate pairs, 
see Fig. \ref{fig:WRMTscat}. 
Moreover, we have found perfect agreement with the analytical predictions
for $\rho_\chi^\nu$ in Eq.~(\ref{rhoWreal}) and $\rho^\nu_5$ in 
Eq.~(\ref{rho5}), this is illustrated by Fig.  \ref{fig:WRMTnumcheck}. 
This agreement strongly suggest that the chiral Random Matrix Theory
for Wilson fermions
introduced above is in the same spectral universality class as 
the leading term of Wilson
chiral Perturbation Theory in the $\epsilon$-counting scheme.

\vspace{2mm}

To demonstrate this universality we will now show that the Wilson chiral 
Random Matrix Theory
partition function reduces to that of Wilson
chiral Perturbation Theory to leading order in the $\epsilon$-counting
scheme.
The explicit calculation will allow us to identify
the low energy constant $W_8$ in terms of the parameters of the 
Random Matrix Theory. Inclusion in the Random Matrix Theory
of the two other terms in the chiral Lagrangian to this order
that couple with $W_6$ and $W_7$ will be discussed in the next
subsection. For notational simplicity we keep the quark flavors
degenerate.

\vspace{3mm}

The first step is to express the determinants in Eq.~(\ref{ZWRMT}) as 
Grassmann integrals. Then the average over the matrix elements of $W$
can be performed by completing squares, see \cite{SV,RMT1}
for details. The four-fermion terms can be decoupled by means of a 
Hubbard-Stratonovitch transformation, leading to the partition function
\be\label{zavex}
\tilde{Z}^\nu_{N_f}(\tm,\tz;\ta) 
& = & \int d Q_1 d Q_2 d T d T^\dagger  \\ 
&&\hspace{-1cm}\times {\det}^{N/2+\nu}(i \ta Q_1 + T +\tm+\tz)
{\det}^{N/2}(i \ta Q_2 + T^\dagger +\tm -\tz)
e^{-\frac N2{\rm Tr} T T^\dagger -\frac{N}{4} {\rm Tr}[Q_1^2 +Q_2^2]}. \nn
\ee
Here, $Q_1$ and $Q_2$ are Hermitian $N_f\times N_f$ matrices and
$T$ is an arbitrary complex $N_f\times N_f$ matrix.

Up to now this is an exact reformulation of the partition function
(\ref{ZWRMT}). 
In the limit of large matrices valid for
the scaling (\ref{scaling}) the integrals can be evaluated by a 
saddle point approximation. The saddle point manifold is given by
\be
T= U V^{-1}.
\ee
After absorbing $V^{-1}$ in $U$ and expanding the exponent
to order $\tm$, $\tz$ and $\ta^2$ we obtain
\be 
{\tilde Z}^\nu_{N_f}(\tm,\tz;\ta) & = & \int_{U(N_f)} 
d U \int dQ_1 dQ_2 \ {\det}^\nu(U) \
e^{ \frac N2 (\tm+\tz){ \rm Tr} U
+ \frac N2  (\tm-\tz){ \rm Tr} U^\dagger}
\nn \\ && \times
e^{ \frac N2  i\ta{\rm Tr}(Q_1U + Q_2 U^{-1}) +
\frac N4 \tilde{a}^2 {\rm Tr}[ Q_1 U^{-1}Q_1U^{-1} + Q_2 UQ_2U]
-\frac N4 {\rm Tr}(Q_1^2 +Q_2^2)
}. \nn
\ee
The integrals over $Q_1$ and $Q_2$ are Gaussian and are localized 
on the saddle point.
The term of $O(\tilde{a}^2)$ in the exponent results in higher order 
contributions and they can be omitted here.
We thus obtain  the partition function
\be
\tilde{Z}^\nu_{N_f}(\tm,\tz;\ta)= \int_{U(N_f)} d U \  
{\det}^\nu(U) \ e^{ \frac N2 (\tm+\tz){ \rm Tr} U
+  \frac N2 (\tm-\tz){ \rm Tr} U^\dagger
-N \frac{\ta^2}{4}{\rm Tr}(U^2 + {U^\dagger}^2).
}
\ee
This shows that the Wilson chiral Random Matrix Theory partition
function (\ref{zavex}) reproduces the three leading terms in the
$\epsilon$-expansion. The corresponding low energy constants are 
identified as
\be
N\tm = {m\Sigma V}, \quad N\tz =
{z\Sigma V}, 
\quad \frac{N\ta^2}{4} = a^2 W_8 V.
\label{match}
\ee
As discussed in section \ref{subsec:WRMTW8} this identification requires that 
$W_8>0$. 
 
\subsection{Double-Trace terms}
\label{subsec:W6W7inWRMT}

The double-trace terms of Wilson chiral Perturbation Theory are
not generated by the Wilson Random Matrix Theory defined in 
Eq.~(\ref{ZWRMT}).
Although such terms have been argued to be
suppressed in large-$N_c$ counting \cite{KL}, we would
nevertheless like to be able to include them. As we will now show,
this can easily be done.

Let the Random Matrix Theory partition function be extended to
\be
\tilde{Z}^\nu_{N_f}(\tm,\tz;\ta_6,\ta_7,\ta_8) 
\equiv \frac{1}{16 \pi \ta_6 \ta_7}\int_{-\infty}^\infty dy_6 dy_7 \ 
e^{-\frac{y_6^2}{16|\ta_6^2|}-\frac{y_7^2}{16|\ta_7^2|}} \ 
\tilde{Z}_{N_f}^\nu(\tm-y_6,\tz-y_7;\ta_8) ,   
\label{zrandmass}
\ee
where the partition function inside the integral is given in (\ref{ZWRMT}).
It follows from the discussion of subsection
\ref{subsec:W6W7inWCPT} that then also the trace-squared terms
of Wilson chiral Perturbation Theory are reproduced by the Random
Matrix Theory in the scaling limit (\ref{scaling}). The partition functions 
(\ref{zrandmass}) lead to 
negative values of $W_6$ and $W_7$. Positive values
of these constants can be obtained from fluctuations of the mass in the 
imaginary direction.

\section{Hermiticity Properties and the Sign of $W_8$}
\label{sec:sign}

The sign and magnitude of the low energy constants $W_6$, $W_7$ and
$W_8$ are essential for numerical simulations of lattice QCD with Wilson
fermions. In particular the sign of $W_8$ has been debated in the literature
\cite{Sharpe2006,Azcoiti:2008dn,Shindler:2009ri}. The sign is
important for understanding whether or not lattice QCD with Wilson
fermions enters an Aoki
phase with spontaneously broken parity \cite{Aokiclassic,Heller}. 
In this section we
provide several arguments for why Hermiticity properties put constraints
on the these low-energy parameters. To simplify the discussion, we
restrict ourselves to the case where both $W_6$ and $W_7$ vanish.

\subsection{A Wilson lattice QCD inequality and the sign of $W_8$}
\label{subsec:QCDineq}

With two degenerate flavors the determinant of the Wilson Dirac
operator is positive definite and it is possible to formulate rigorous 
QCD inequalities. Based on such an inequality we argue here 
that for $W_6=W_7=0$ one must have $W_8\geq0$ in the $N_f=2$ theory. 
Since the Wilson fermion determinant in lattice QCD is real, the
measure with two degenerate flavors is real and positive
\be
{\det}^2(D_W+m) > 0. 
\ee
As this is true for any gauge field 
configuration, independent of the number of
real eigenvalues of $D_W$, it follows that the two-flavor partition
function, ${\cal Z}^\nu_{N_f=2}$, of lattice QCD with Wilson fermions 
in a sector with fixed number of real modes of $D_W$  
is real and positive for all real values of $m$
\be
{\cal Z}^\nu_{N_f=2}(m,z=0;a)  > 0.
\label{QCDineq}
\ee
The overall sign of the partition function can of course be changed by
introducing a multiplicative constant. What the inequality states
is that the partition function cannot change sign as a function of 
a real valued quark mass.  

The partition function in the $\epsilon$-regime must necessarily 
satisfy the same positivity bound. We note, however, that its sign
depends on the index $\nu$ of the chiral Lagrangian:
\be
Z_{N_f=2}^\nu(\hm,\hz;-\ha_8^2) = (-1)^\nu 
Z_{N_f=2}^\nu(i\hz,i\hm;\ha_8^2).
\ee\begin{center}
\begin{figure}
\includegraphics[width=8.5cm,angle=0]{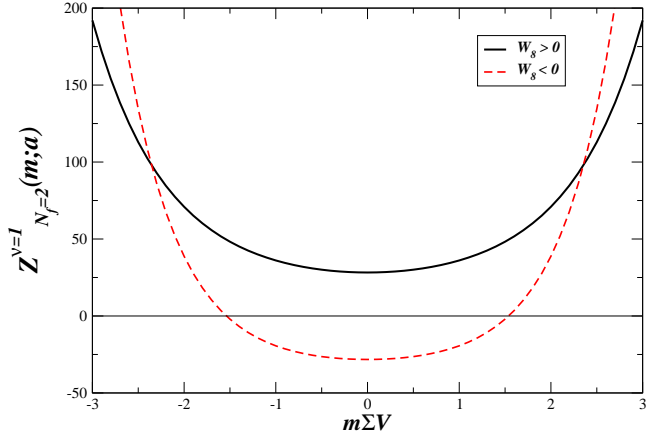}
\caption{\label{fig:QCDineq} The two flavor microscopic partition function for
  $\ha_8^2=1$ ($W_8\geq0$) and $\ha_8^2=-1$ ($W_8 \leq0$) plotted as a
  function of the mass for $\hz=0$. Only the partition function with $W_8\geq0$
  satisfies the QCD inequality.}
\end{figure}
\end{center}
For odd $\nu$ the partition functions with positive
and negative $\ha_8^2$ have opposite signs at $m=z=0$ and, since for
sufficiently large $\hm$ both partition functions have the same sign,
it is not possible that they both satisfy the inequality for real
$\hm$ and $\hz$. As the plot in  
Fig. \ref{fig:QCDineq} demonstrates, the sign of $W_8$ dictated 
by the inequality is positive. 

This can also can be obtained analytically from the
chiral Lagrangian. The partition function with degenerate masses 
(\ref{ZLfull}) can be expressed in terms of the eigenvalues of
$U$. This allows us to rewrite the mass-degenerate partition function
in the simplified form    
\be
Z_{N_f}^\nu(\hm,\hz;\ha_8) & = &
\det[Z^{\nu+i-j}_{N_f=1}(\hm,\hz;\ha_8)]_{i,j=1\ldots N_f} 
\label{ZNf}
\ee
where the one flavor partition function is 
\be
\label{ZNf1}
Z_{N_f=1}^\nu(\hm,\hz;\ha_8) & = & 
\int_{-\pi}^\pi \frac{d\theta}{2\pi} \ e^{i\theta\nu} 
e^{\hm\cos(\theta)+i\hz\sin(\theta)-2\ha_8^2\cos(2\theta)}. 
\ee
In \ref{app:C} we show that for odd values of $\nu$ 
the two-flavor partition function with $W_8>0$ ($W_8<0$) is positive
(negative) for $\hm=0$ and that for both signs of $W_8$ the curvature
at zero mass is positive.

Below we discuss the sign of $W_8$ from the perspective of the graded 
generating function and from Wilson chiral Random Matrix Theory.
In both instances we find $W_8\geq0$, independent of the number of
flavors.

\subsection{Convergence and $\gamma_5$-Hermiticity}
\label{subsec:conv}

The issue of the sign of the constants of Wilson
chiral Perturbation Theory manifests itself also in the convergence
properties of the graded generating functional. 
For $W_6=W_7=0$ the integrals in (\ref{ZSUSY}) are divergent
if $W_8 < 0$. However, the alternative graded generating function  
\be
\label{ZSUSYord}
Z_{1|1}^\nu({\cal M},{\cal Z};\ha_8^2<0)= \int \hspace{-1.5mm} dU \
{\rm Sdet}(U)^\nu \ 
  e^{\frac{1}{2}{\Str}({\cal M}[U+U^{-1}])
    +\frac{1}{2}{\Str}({\cal Z}[U-U^{-1}])
    -\ha_8^2{{\Str}(U^2+U^{-2})}} 
\ee
is now convergent. This integral seems to provide a 
{\em bona fide} generating function for Wilson fermions with
$W_8 < 0$. In fact, it agrees precisely with the generating function
suggested for $p$-regime calculations in ref. \cite{Sharpe2006}. 
Nevertheless, the sign-issue has not disappeared, it has only resurfaced in
disguise. To see this, 
we can compute the resolvent for the real eigenvalues of
$D_W$ with the alternative graded generating function

\be
\left . \frac d{dm} Z_{1|1}^\nu({\cal M},{\cal Z}=0;\ha_8^2<0)\right |_{m'=m}.
\ee
By comparison with Eqs. 
(\ref{ZSUSY}) and (\ref{ZSUSYord}), it follows 
that
\be
\left. \frac d{dm} Z_{1|1}^\nu({\cal M},{\cal Z}=0;\ha_8^2<0)
\right|_{m'=m}
= -i\left . \frac d{dz} Z_{1|1}^\nu({\cal M}=0 ,{\cal Z};\ha_8^2>0)
\right|_{z'=z=im}.
\ee
Since the rhs is continuous when $m$ crosses the real axis
we conclude that for $W_8<0$ the density of real
eigenvalues vanishes. This is because the graded partition
function, (\ref{ZSUSYord}), valid for $W_8<0$, is the generating
functional of an anti-Hermitian operator with a spectral density 
on the imaginary axis. 
This would be an acceptable eigenvalue density if $D_W$ was
anti-Hermitian, but this is only the case for $a=0$. Away from
$a=0$ the operator is only $\gamma_5$-Hermitian. So from the
convergence of the graded generating function and the
requirement of
$\gamma_5$-Hermiticity of $D_W$ it follows that $W_8>0$.

\vspace{2mm}

The possibility of a link between the {integration domain} 
of the graded generating functional in chiral Perturbation Theory 
and the Hermiticity properties of the Dirac operator is not new. 
In continuum QCD at non-zero quark chemical potential the Dirac operator 
is non-Hermitian for real values of the chemical potential and 
anti-Hermitian for imaginary values. Correspondingly, the graded
generating functional has two branches which differ by a
transformation  
of the super Goldstone field \cite{Factorization,Bosonic}. One is 
convergent for real values of the chemical potential while the other 
is convergent for imaginary values of the chemical potential
\cite{Fpi1,Fpi2}.

\subsection{The sign of $W_8$ and Wilson chiral Random Matrix Theory}
\label{subsec:WRMTW8}

In this section we discuss the constraints on the constants in Wilson chiral
Perturbation Theory due to $\gamma_5$-Hermiticity from the point 
of view of chiral Random Matrix Theory.

In Eq. (\ref{match}) the coefficients of Wilson chiral Perturbation Theory
where expressed in terms of the parameters of the chiral Random Matrix
Theory, and it was found that $W_8 >0$.
A chiral Random Matrix Theory where the sign of
$W_8$ is negative can be obtained by including an additional 
imaginary unit $i$ in front of the chiral symmetry violating pieces 
of $\tilde{D}_W$ in 
(\ref{w-diracOP}), that is 
\be
\tilde{D}_{W_8<0} = \mat iaA & iW \\ iW^\dagger & iaB \emat.
\label{DW8neq}
\ee
With this change, the Random Matrix Wilson Dirac
operator, however, becomes anti-Hermitian rather than 
$\gamma_5$-Hermitian. 
We stress that this observation is in perfect agreement with the 
conclusion obtained from the convergence requirement of the  
graded generating function above: the spectral density 
of the anti-Hermitian Random Matrix (\ref{DW8neq}) matches 
that generated by the graded partition function 
(\ref{ZSUSYord}) which is convergent for $W_8<0$. This has been
verified numerically to high accuracy. 
Because $\tilde{D}_{W_8<0}$ is anti-Hermitian, the determinant
$\det(\tilde{D}_{W_8<0}+m)^2$ has a definite sign for imaginary $m$ 
in agreement with the chiral Lagrangian for $W_8 < 0$ and imaginary $m$.

From the point of view of Random Matrix Theory, the different
universality classes corresponding to whether one imposes
$\gamma_5$-Hermiticity or not are well understood \cite{Magnea}. 
In the Random Matrix Theory literature this is referred to
as $Q$-symmetry.

\vspace{3mm}

Clearly the conclusions of this section have the same origin, 
namely that the case $W_8<0$ (and $W_6=W_7=0$) is in conflict with the 
$\gamma_5$-Hermiticity of $D_W$ and the Hermiticity of $D_5$. 
A positive sign of $W_8$ is consistent with the existence of an
Aoki phase.

\section{Conclusions}

In this paper we analyzed the  spectral properties of the Wilson Dirac
operator in the microscopic scaling regime. Eigenvalues of the
Wilson Dirac operator in this regime are responsible for
spontaneous chiral symmetry breaking. Another feature of 
Wilson fermions is the existence of the
Aoki phase with spontaneous breaking of parity.
In the microscopic scaling regime one has analytical
means for studying both. We have shown that the chiral
Lagrangian for Wilson Chiral Perturbation Theory can be
used to compute spectral properties of the Wilson Dirac
operator $D_W$ and its Hermitian counterpart $D_5$. 
Using a graded extension of the chiral Lagrangian up to
and including ${\cal O}(a^2)$ effects, we have computed the
microscopic spectrum of the Hermitian Wilson Dirac operator
$D_5$ in the quenched theory. We have shown how to incorporate
all possible terms of the Lagrangian to this order, including
double trace terms.

An alternative path to the microscopic spectrum goes through 
a chiral Random Matrix Theory based on symmetries of the
Wilson Dirac operator. Here we have formulated such a theory and 
proved that the partition function coincides with that of Wilson
chiral Perturbation Theory to leading order in the $\epsilon$ 
counting scheme. 
Numerically, we have demonstrated that 
also the spectral correlation functions of the chiral Random Matrix
Theory agree beautifully with the analytical predictions of Wilson
chiral Perturbation Theory.

Interestingly, we find 
restrictions on the possible values of the low-energy
constants of Wilson Chiral Perturbation Theory arising from the
imposition of $\gamma_5$ Hermiticity. It would
be most interesting to have a more detailed understanding of
this phenomenon, which seems to run deep. 
It is noteworthy that the bound we get
from the chiral Lagrangian matches exactly with the bound
we get, by an entirely different route, from our Wilson chiral
Random Matrix Theory. In its simplest form, where we ignore
double trace terms, it was found that this bound is consistent with the
existence of an Aoki phase. 

The results presented here are are for the microscopic limit in 
sectors with a fixed index of the Wilson Dirac operator. 
The real modes of the Wilson Dirac operator have an interesting
dynamics on their own. Remarkably, we find that 
very close to the continuum, the $|\nu|$  
real modes behave like the eigenvalues of a $|\nu|\times|\nu|$ 
Hermitian Random Matrix Theory of exactly Gaussian weight. 
We emphasize that this is a universal result that has been derived
from a chiral Lagrangian. In lattice gauge
theory simulations it would be most interesting to study just
this subset of real modes of the Wilson Dirac operator.
In particular, as we have shown, their dynamics carries detailed
information about the
low-energy constants of Wilson Chiral Perturbation Theory.

In general, we suggest to use spectral properties of the
Wilson Dirac operator to determine the new parameters of
Wilson Chiral Perturbation Theory. While these parameters
are unphysical, it is nevertheless crucial to establish
their values if one wishes to extract physical observables
from Wilson fermion simulations at finite lattice spacings.
It will thus be of obvious interest to extend the first results
presented here for the quenched theory to the corresponding theories
with $N_f$ light fermions. Work in this direction is
presently underway \cite{latNf1}.

\vspace{2mm}

\noindent
{\bf Acknowledgments:}
We would like to thank many participants of the Lattice 2010 Symposium and
the TH-Institute ``Future directions in lattice
gauge theory - LGT10'' at CERN for discussions. We thank C. Gattringer
for clarifying correspondence. Two of us (GA and JV) would
like to thank the Niels Bohr Foundation for support and the Niels
Bohr Institute and the Niels Bohr International Academy for its
warm hospitality. This work was supported  by U.S. DOE Grant No. 
DE-FG-88ER40388 (JV) and the 
Danish Natural Science Research Council (KS). 

\newpage

\renewcommand{\thesection}{Appendix \Alph{section}}
\setcounter{section}{0}

\section{Small $\ha_8$-Limit of Resolvent for $(\hz-\hm)/\ha_8$ Fixed.}
\label{app:A}

In this appendix we consider the leading nontrivial expansion of
$\rho_5^\nu$ for small $\ha_8$ at $\nu\geq0$. We start from
(\ref{res-result}) with $W_6=W_7=0$ and then compute the small
$\ha_8$ limit for $(\hz-\hm)/\ha_8$ fixed. Our aim is to prove that in
this limit the eigenvalue 
density is given by the familiar expression for the density of the
Gaussian Unitary Ensemble with matrices of size $\nu\times\nu$.

\vspace{3mm}

The fermionic $\theta$-integrals in (\ref{res-result}) are all of the form
\be
Z_p^f &\equiv &\frac 1{2\pi} \int_{-\pi}^\pi d\theta e^{i p\theta} e^{-\hm\sin\theta +i\hz\cos\theta +2\ha_8^2 \cos 2\theta},\ \ \ \mbox{for}\ \ p=\nu-3,\ldots,\nu+3\nn \\
&= &
\frac 1{2\pi} \int_{-\pi}^\pi d\theta e^{i p\theta} e^{\sqrt{\hm^2 - \hz^2}
\cos(\theta+\phi)  +2\ha_8^2 \cos 2\theta} \nn \\
&=&
\frac 1{2\pi} \int_{-\pi}^\pi d\theta e^{i p(\theta-\phi)} e^{\sqrt{\hm^2 - \hz^2}
\cos(\theta)  +2\ha_8^2 \cos 2(\theta-\phi)},
\ee
with
\be
e^{-i\phi} = \left ( \frac {\hz-\hm}{\hz+\hm}\right )^{1/2}\qquad {\rm and } \qquad
\cos \phi = \frac {i\hz}{\sqrt{\hm^2-\hz^2}},\qquad
 \sin \phi = \frac \hm {\sqrt {\hm^2-\hz^2}}.
\label{phidef} 
\ee
To obtain the small $\ha_8$  and small $\hm-\hz$ limit we expand the
exponential functions in a  Taylor series
\be
Z_p^f &=&
\frac 1{2\pi} \int_{-\pi}^\pi d\theta e^{i p(\theta-\phi)}
\sum_{k,l =0}^\infty  \frac 1{k!l!} 
(\hm^2-\hz^2)^{k/2} \cos^k\theta (2 \ha_8^2 \cos 2(\theta-\phi))^{l}.
\ee
The leading terms, which we will refer to as $\check Z_p^f$, are
obtained from exponents in the cosines with opposite sign to the sign
of $p$ 
\be
\check Z_p^f & \equiv &
 e^{-i p \phi}
{\sum_{k+2l = |p|}}  \frac 1{k!l!2^k}
(\hm^2-\hz^2)^{k/2}  \ha_8^{2l} e^{ 2i l\, {\rm sign}(p) \phi} \nn\\
&=&
{\sum_{k+2l = |p|}}  \frac 1{k!l!2^k}
(\hm^2-\hz^2)^{k/2}  \ha_8^{2l} e^{ -i k\, {\rm sign}(p) \phi}\nn\\
&=&
(\ha_8\sqrt 2)^{|p|}{\sum_{l=0}^{[|p|/2]}}  \frac {\alpha^{|p|-2l}}{(|p|-2l)!l!2^l}, 
\label{zfer1}
\ee
where we define
\be
\alpha = \frac {i(\hz-{\rm sign}(p) \hm)}{2\ha_8\sqrt 2}.
\label{alpha}
\ee
For $p=0$ we have $\check Z_{p=0}^f=1 $.
The sum above can be expressed in terms of Hermite polynomials,
normalized according to (\ref{normH}), for all $p$,
\be
\check Z_p^f =(\ha_8\sqrt 2)^{|p|} \frac{i^{|p|}}{2^{|p|/2}\ |p|!} H_{|p|}(\alpha/(i\sqrt 2)),
\label{ZHrel}
\ee
where the only dependence on the sign of $p$ is in the argument 
$\alpha$ (noting that $H_0=1$).

We now turn to the bosonic $s$-integrals in (\ref{res-result}). They are all of the form
\be
Z_p^b&\equiv& \int_{-\infty}^\infty ds \ e^{-p
  s}e^{-i\hm\sinh(s)-i\hz\cosh(s)-2\ha_8^2\cosh(2s)}\ , \ \ \ \mbox{for}\ \
p=\nu-2,\ldots,\nu+2\ ,\nn \\ 
&=& \int_0^\infty \frac{dy}{y} \ y^{-p}e^{-\frac{i}{2}(\hm +\hz)y
  -\frac{i}{2}(\hz -\hm) 
/y-\ha_8^2(y^2+y^{-2})},
\label{Zpbexact}
\ee
after using the same definition as in (\ref{phidef}) and changing variables $y=e^s$.
In the limit
$a\to 0$ with $|\hm-\hz|/\ha_8$ fixed the leading contribution in an expansion in powers of
$\ha_8$ is given as follows. 

For $p>0$ ($p<0$) we change to rescaled new variables
$y=\ha_8\sqrt{2}/t$ ($y=t/\ha_8\sqrt{2}$) to obtain to leading order 
\be
\check Z_p^b
&\equiv &(\ha_8\sqrt{2})^{-|p|}\int_0^\infty dt \ t^{|p|-1} \exp[-\alpha t-t^2/2]\nn\\
&=& (|p|-1)!D_{-|p|}(\alpha)
\frac{e^{\alpha^2/4}}{(\ha_8\sqrt 2)^{|p|}},
\ee
with $D_p$ a parabolic cylinder function \cite{Abramowitz}, and $\alpha$ 
depending on sign$(p)$ defined in (\ref{alpha}).

For $p=0$ the behavior in $\ha_8$ is different. Because of the exact result 
$Z_p^b(\hz=0=\hm)=K_0(2\ha_8^2)$ we expect a logarithmic singularity here. 
Splitting the integration in Eq. (\ref{Zpbexact}) into $(0,1)$ and
$(1,\infty)$ and changing variables as for $p>0$ for the former integral, and
as for $p<0$ for the latter, we arrive at
\be
Z_{p=0}^b&=& \int_{\ha_8\sqrt{2}}^\infty\frac{dt}{t}\left[ 
e^{-\frac{i}{2}(\hm+\hz)\frac{\ha_8\sqrt{2}}{t} -\frac{i}{2}(\hz-\hm)
\frac{t}{\ha_8\sqrt{2}}}
\right.\nn\\
&&\hspace{2cm}+\left. 
e^{-\frac{i}{2}(\hm+\hz)\frac{t}{\ha_8\sqrt{2}} -\frac{i}{2}(\hz-\hm)
\frac{\ha_8\sqrt{2}}{t}}
\right]e^{-\ha_8^2(2\ha_8^2/t^2+t^2/(2\ha_8^2))}.
\ee
Because the saddle point is outside the integration contour the leading
contribution comes from the lower endpoint of integration $\ha_8\sqrt{2}$, and
we arrive at
\be
\check Z_0^b=-\log(2\ha_8^2)\exp[-i(\hm+\hz)/2-i(\hz-\hm)/2].
\label{Zp0b}
\ee

By combining the fermionic and bosonic integrals we obtain 
\be
G^\nu(\hz,\hm,\ha_8)&\equiv &
\frac i2\Big[ Z_\nu^b\left\{
\frac i4(\hm+\hz)[ Z_{\nu+2}^f+ Z_{\nu}^f]
+\frac i4(\hz-\hm)[ Z_{\nu}^f+ Z_{\nu-2}^f]
+\frac12[ Z_{\nu+1}^f+ Z_{\nu-1}^f] \right.\nn\\
&&\left.
\ \ \ \ \ \ \ \ +\ha_8^2
[ Z_{\nu+3}^f+ Z_{\nu+1}^f+ Z_{\nu-1}^f+ Z_{\nu-3}^f]
\right\}\nn\\
&&+( Z_{\nu+1}^f+ Z_{\nu-1}^f)\left\{ 
\frac i4(\hm+\hz) Z_{\nu-1}^b+\frac i4(\hz-\hm) Z_{\nu+1}^b
+\ha_8^2 [ Z_{\nu-2}^b+ Z_{\nu+2}^b] 
\right\} \nn\\
&&+2\ha_8^2 Z_{\nu-1}^b [ Z_{\nu+2}^f+ Z_{\nu}^f]
+2\ha_8^2  Z_{\nu+1}^b [ Z_{\nu-2}^f+ Z_{\nu}^f]\Big]\ .
\label{Gfull}
\ee
To obtain the leading small $\ha_8$ limit we simply replace the
partition functions by those with a caron. 

For $\nu \ge 3$ all bosonic integrals have a positive subscript $p>0$
and no log-terms appear. 
For the fermionic integrals for  $\nu \ge 3$
it is immediately clear that in the limit $\ha_8 \to 0$ only
the fermionic partition functions with the
lowest index have to be taken into account in each sum.
The cases $\nu =0,1,2$ have to be checked separately.  
Here special attention has to be paid to the fact that for sign$(p)<0$ 
we have $\alpha\sim 1/\ha_8$. 

On the fermionic side 
the term $\check Z_{\nu+3}^f+\check Z_{\nu+1}^f+\check Z_{\nu-1}^f+\check
Z_{\nu-3}^f$ is subleading both for $\nu =1 $ and $\nu =2$ whereas 
$\check Z_{\nu}^f+\check Z^f_{\nu-2}$ is subleading for $\nu = 1$. (This is
also the case for 
$z=0$ relevant for the resolvent of the real eigenvalues of $D_W$.) 
Below this is automatically taken care of by the factors $(\nu-1)$ and
($\nu-2$). 

For $\nu>0$ 
the expression for the resolvent thus simplifies to leading order to 
\be
\label{GLO}
&&G^{\nu>0}(\hz,\hm;\ha_8)=\\
&&\frac i2 \frac {i^{\nu-1}e^{\alpha^2/4}}{2^{(\nu-1)/2}\ha_82\sqrt 2} \left [
H_{\nu-1}(\frac{\alpha}{i\sqrt{2}})
[D_{-\nu}(\alpha)+\nu \alpha D_{-1-\nu}(\alpha)+\nu(\nu+1)D_{-\nu-2}(\alpha)]
\right . \nn \\ && \left .
+\frac{\sqrt{2}(\nu-1)}{i}H_{\nu-2}(\frac{\alpha}{i\sqrt{2}}) [\alpha
  D_{-\nu}(\alpha)+2\nu D_{-\nu-1}(\alpha)] 
+\frac{2(\nu-1)(\nu-2)}{i^2}H_{\nu-3}
(\frac{\alpha}{i\sqrt{2}})D_{-\nu}(\alpha) 
\right ].\nn
\ee
For $\nu=0$ we obtain no order $1/\ha_8$ terms, and the leading order is in
this particular case given by 
\be
\frac 18\log(2\ha_8^2)(\hz+\hm) 
\exp[-i(\hm+\hz)/2].
\ee

Using the recursion relation 
\be
D_{p+1}(\hz) -\hz D_p(\hz) +pD_{p-1}(\hz)= 0,
\label{Drec}
\ee
one can simplify the first term in Eq. (\ref{GLO}) 
 \be
&& G^{\nu>0}(\hz,\hm;\ha_8)\\ &=&
\frac {i^{\nu}e^{\alpha^2/4}}{2^{(\nu+3)/2}\ha_8\sqrt 2}\left [
(\nu+1)H_{\nu-1}(\frac{\alpha}{i\sqrt{2}})
D_{-\nu}(\alpha)
\right . \nn \\ && \left .
-i{\sqrt{2}(\nu-1)}H_{\nu-2}(\frac{\alpha}{i\sqrt{2}}) [\alpha
  D_{-\nu}(\alpha)+2\nu D_{-\nu-1}(\alpha)] 
-{2(\nu-1)(\nu-2)}
H_{\nu-3} (\frac{\alpha}{i\sqrt{2}})D_{-\nu}(\alpha) 
\right ] . \nn
\ee
For $p$ a positive integer, the parabolic cylinder functions can be written as
\be
D_{-p}(\alpha)= \sqrt{\frac\pi2}\frac{(-i)^{p-1}}{2^{(p-1)/2}(p-1)!}
H_{p-1}(\alpha/(i\sqrt 2))e^{\alpha^2/4}{\rm erfc}(\alpha/\sqrt{2}) +
P_{p-2}(\alpha)e^{-\alpha^2/4}, 
\label{Derfc}
\ee
where we define the following polynomials that have parity $k$
\be
P_k(x)= \sum _{l=0}^{[k/2]}a_l x^{k-2l}
\ee
with real coefficients $a_l$. This relation follows from induction, using the recurrence relation (\ref{Drec}) as well as the following \cite{Abramowitz}
\be
D_n(x)&=&2^{-n/2}e^{-x^2/4}H_n(x/\sqrt{2})\ ,\ \ \mbox{for}\ n=0,1,2,\ldots\nn\\
D_{-1}(x)&=&\sqrt{\frac\pi2}e^{x^2/4}{\rm erfc}(x/\sqrt{2})\ . 
\ee
By inspection it is clear that the terms containing the polynomial $P_k$
do not contribute to the imaginary part of the resolvent. Because of
erfc$(x)=1-$erf$(x)$ and erf$(x)$ being odd,  
only the term proportional to unity contributes to the imaginary part.
After using the recursion relation for the
Hermite polynomials, 
\be
H_{n+1}(x)=2xH_n(x)-2nH_{n-1}(x)\ ,
\ee
we find for the imaginary part of the resolvent
and hence the eigenvalue density of $D_5$ (with $\ha_8\to0$ and
$|\hz-\hm|/\ha_8$ fixed valid for $\nu>0$)
\be
\rho_5^{\nu>0}(\hz,\hm;\ha_8) 
&\approx& \frac1\pi{\rm Im}[G^{\nu>0}(\hz,\hm;\ha_8)] \\
&=&
\frac { e^{\alpha^2/2} }{\sqrt{\pi}\ha_82^{\nu+2}(\nu-1)!}
[H_\nu^2(\alpha/(i\sqrt 2)) -H_{\nu+1}(\alpha/(i\sqrt
  2))H_{\nu-1}(\alpha/(i\sqrt 2))].\nn 
\ee
This is the familiar density of the $\nu\times\nu$ Gaussian Unitary
Ensemble shifted by $\hm$ and rescaled by $1/4\ha_8$. 
For $\nu=0$ the density is simply zero at the same order.

 \newpage

\section{Mean Field Limit}
 \label{app:B}
In this Appendix we give more details for the mean field results discussed
in section \ref{mft} (see also \cite{Lamacraft-Simons,Golterman:2005ie}).

\subsection{Mean Field Analysis of Graded Partition Function}
  
For large $mN$, $zN$ and $a^2N$, the integrals in the expression for
the generating function of the 
resolvent can be evaluated by a saddle point approximation. Unless
the Grassmann integrals vanish at the saddle point (we will see below that this
indeed may happen), the  leading
order result can be obtained by putting
 the Grassmann variables equal to zero so that the
integral factorizes into a compact and a noncompact integral which each
can be approximated by a saddle point integral. 

A priori we can choose
either the compact or the noncompact integral to derive the mean field
result for the resolvent.
This limit corresponds to the physical limit of taking the thermodynamic
limit at fixed lattice spacing. 

The mean field approximation to the supersymmetric generating function is
thus given by (provided that the Grassmann integrals do not vanish
at the saddle point).
\be
Z(\hm,\hz,\hz';\ha_8) = Z^{f}(\hat m, \hat z;\ha_8) Z^{b}(\hat m, \hat z';\ha_8)
\ee
resulting in the spectral density
\be
\rho_5(\hat z,\hat m;\ha_8) = {\rm Im} \left[Z^{f}(\hm,\hz)\left . \frac 1{\pi  }\frac d{d\hz'} Z^{b} (\hm,\hz')\right] \right |_{\hz'=\hz}.
\ee

Let us start with the fermionic integral given by
\be
Z^{f}(\hz, \hm; \ha_8) = \int_{-\pi}^\pi\frac {d\theta }\pi  e^{-\hm \sin \theta
+i\hz\cos \theta+2 \ha_8^2 \cos(2\theta)}.\label{zfer}
\ee
For real $\hz$, the imaginary part of the integral vanishes so that
the spectral density is given by
\be
\rho_5(\hat z,\hat m;\ha_8) = Z^{f}(\hz, \hm; \ha_8)  \left . \frac 1{\pi }\frac d{d\hz'} {\rm Im }
Z^{b}(\hz', \hm; \ha_8) \right |_{\hz'=\hz}.
\ee
The fermionic integral can be rewritten as
\be
 Z^{f}(\hz, \hm; \ha_8) = \frac 2\pi {\rm Re}\, \int_{-1}^1\frac {dy}{\sqrt{1-y^2} }  e^{-\hm y
+i\hz\sqrt{1-y^2}+2 \ha_8^2 (1-2y^2)}.
\ee
The extremum of the real part of the exponent is at $y = -\hm/8\ha_8^2$, so
that for $8\ha_8^2 < \hm$, the integral can be approximated by expanding 
about $y = -1$. In the thermodynamic limit we arrive at
\be
Z^{f}(\hz, \hm; \ha_8) &=& \frac{2\sqrt 2} \pi{\rm Re}\, \int_0^\infty  ds e^{\hm-2\ha_8^2-(\hm-8\ha_8^2)s^2+i\hz s\sqrt 2}\nn \\ 
&=&
\sqrt{\frac 2\pi}\frac {e^{\hm-2\ha_8^2-\hz^2/2(\hm-8a^2)}}{\sqrt{\hm-8\ha_8^2}}.\label{zferasym}
\ee
The $z$-dependence in the exponent is subleading in the above
expression, so that
the fermionic part of the partition function does not contribute to the
resolvent to leading order.
For $ \ha_8= 0$ the integral in  Eq. (\ref{zfer}) is given by (see Eq. (\ref{iint}))
\be
Z^{f}(\hz,\hm; \ha_8 = 0) = 2 I_0(\sqrt{\hm^2-\hz^2}) \sim\frac{e^{\sqrt{\hm^2-\hz^2}}}
{2 \sqrt{2\pi} (\hm^2-\hz^2)^{1/4}}. 
\ee
For $\hz \ll \hm$ this expression is approximated by
\be
\sqrt{\frac 2 \pi} \frac{e^{\hm-\hz^2/2\hm}}{\sqrt{\hm}}.
\ee
in agreement with the asymptotic result (\ref{zferasym}) for $ a =0 $.

Next we consider the bosonic integral given by
\be
Z^{b}(\hz,\hm; \ha_8) = \int_{-\infty}^\infty ds e^{-i\hm\sinh s -i\hz \cosh s-2\ha_8^2\cosh 2s}.
\ee
The imaginary part of the partition function can be written as
\be
{\rm Im}
Z^{b}(\hz,\hm; \ha_8) = \frac 1{2i}\int_{-\infty}^\infty ds 
[e^{i\hm\sinh s +i\hz \cosh s-2\ha_8^2\cosh 2s}
-e^{-i\hm\sinh s -i\hz \cosh s-2\ha_8^2\cosh 2s}].
\ee
It is convenient to shift
the integration contour by $-\pi i /2$ so that the exponent becomes real. 
The saddle point approximation to the bosonic integral is then given by
\be
Z^{b \ {\rm sp}}(\hz,\hm; \ha_8)  =  \sum_p \left ( \frac { \pi }{  -S_{\rm s}''(s_p)}\right )^{1/2} e^{S_{\rm s}(s_p)},
\ee
where 
\be
S_s(s) = -\hm\cosh s -\hz \sinh s+2\ha_8^2\cosh 2s, \label{saddlepot}
\ee
and the sum is over the saddle points. For $\hz <\hz_g$ the saddle points are real
and only the saddle points with $ S_{\rm s}''(s_p)> 0$ 
contribute to $ {\rm Im}(Z^{b}) $.
There is one saddle point with $ S_{\rm s}''(s_p)< 0$ which gives the
real part of $Z^{b} $ 
and is of the order of $(Z^{f})^{-1}$ so that the graded partition 
function is normalized correctly at the mean field level. We thus find
that for $z < z_g$ the real and imaginary parts of $Z^{b} $ are
determined by different saddle points.

At $z=z_g$ the real saddle point with negative curvature merges with the real
saddle point with positive curvature (which determines the imaginary part of
the partition function), and for $z>z_g$ they turn into a pair of complex
conjugate saddle points. However, only one of the two saddle points is
accessible resulting in a bosonic partition function with a real and an imaginary part
that are both determined by the same saddle point. 

Since the negative of the fermionic exponent is obtained from the bosonic
exponent by replacing $s \to i \theta$, the saddle points
for the bosonic and fermionic integral are the same but in the fermionic
case both saddle points of the complex conjugate pair contribute to the
partition function resulting in a real expression.

Notice that replica symmetry or supersymmetry is broken for the contribution
to the tail. The fermionic integral is always real and the imaginary part
is due to the bosonic integral \cite{martin-critique}.
The resolvent that can be derived form the fermionic partition function does
not have an imaginary part for any number of flavors. 
The replica trick therefore
fails for the fermionic partition function even at the mean field level.
To get the correct result we have 
to select one of the two saddle points. This is the case for
 the bosonic partition function where only
one of the two saddle points is accessible by deformation 
of the integration contour. 

A leading order saddle point approximation for the imaginary part of the  
bosonic partition function is accurate for a large parameter range.
In particular, for large $\hat m$ and $\hz$ it covers both large and
small $\hat{a}_8$. 
There are two parameter domains where the derivation simplifies. First, for $(\hm-\hz)/\ha_8 \gg 1$, then
$\cosh s$ in Eq. (\ref{saddlepot}) can be approximated by $\sinh(s)$. Second, for $\hz$ close to the edge of the spectrum,
where two real saddle points are close and the potential (\ref{saddlepot}) can be approximated by
a cubic potential. We first discuss the small $\ha_8$ limit.

\subsection{Tail of Eigenvalue Distribution for $(\hm-\hz)/\ha_8 \gg 1$.}
 
It is convenient to introduce $u=\sinh s$ as new variable so  the potential
(\ref{saddlepot}) is given by Eq. (\ref{saddleu}).
For $(\hm-\hz)/\ha_8 \gg 1$, the leading saddle point $\bar u \gg 1$, so that
$\sqrt {1+u^2} \approx |u|$. For $ \hz> 0$ the leading saddle point is negative
(and mutates mutandis for $\hz< 0$), so that
\be
S_b(u) \approx  \hm u - \hz u +2\ha_8^2 (2u^2+1), \qquad \hz >
0. \label{saddlesmalla} 
\ee
Taking into account the Jacobian of the transformation $u=\sinh s$ we arrive at
\be
{\rm Im} Z^{b}(\hz,\hm; \ha_8) \sim 4 \sqrt \pi \frac{\ha_8}{\hm-\hz} 
e^{-(\hm-\hz)^2/16\ha_8^2+2\ha_8^2}.
\ee
Combining this with the fermionic integral we obtain for the spectral density
\be
\rho_5(\hz,\hm; \ha_8) = \frac {4\sqrt 2}\pi \frac{\ha_8}{(\hm-\hz)
\sqrt{\hm-8\ha_8^2}}e^{\hm-\hm^2/2(\hm-8\ha_8^2) -(\hm-\hz)^2/16\ha_8^2}.
\ee

\subsection{Edge Scaling}

In this section we study the Dirac spectrum near the gap at  $\hz = \hz_g$.
We first consider $\hz< \hz_g$ and then analyze $\hz> \hz_g$.

\subsubsection{Edge Scaling for $\hz<\hz_g$}

We now expand $S_b(u)$ near the edge of the spectrum. The second derivative
vanishes and the first derivative is given by
\be
S_b'(u_g) = \hz_g- \hz
\ee
resulting in the expansion
\be
S_b(u) = S_b(u_g)-(u-u_g)(\hz-\hz_g) + \frac 16 (u-u_g)^3 S'''_b(u_g)
\ee
with
\be
S'''_b(u_g) = -\frac{3\hm u_g}{(1+u_g^2)^{5/2}}.
\ee
The saddle points are given by 
\be
\bar u = u_g \pm \left ( \frac{2(\hz-\hz_g)}{S'''(u_g)}\right )^{1/2}.
\label{saddlepoints}
\ee
We will see that the minus sign corresponds to the leading saddle point
of the imaginary part of the bosonic partition function whereas the 
negative sign corresponds to the leading saddle point of the fermionic 
partition function as well as the real part of the bosonic partition function.
 Notice, that  because of the supertrace, the fermionic
and bosonic actions are each others inverses.
The bosonic exponent at the saddle point for the imaginary part becomes
\be
S_b(\bar u) = S_b(u_g) - \frac 13 \frac {(2(\hz_g-\hz))^{3/2}}
{\sqrt{S_b'''(u_g)}},
\ee
whereas the fermionic exponent at the saddle point reads
\be
S_b(\bar u) = -S_b(u_g) - \frac 13 \frac {(2(\hz_g-\hz))^{3/2}}
{\sqrt{S_b'''(u_g)}}.
\ee

The second derivative at the saddle points is given by
\be
S''(\bar u) &=& S''(u_g) + (\bar u -\bar u_g) S'''(u_g),\nn\\
&=&  (\bar u -\bar u_g) S'''(u_g),
\ee
so that 
\be
S_b(u) &=& S_b(\bar u) +\frac 12 (u -\bar u)^2(\bar u - u_g) S'''_b(u_g)\nn \\
&=&S_b(\bar u_g) +\frac 12(u-\bar u)^2 [2(\hz-\hz_g)S'''(u_g)]^{1/2}.
\ee
Therefore, integration over $u$ gives an  overall factor $i$ for $\hz < \hz_g$ 
for  the  saddle point $\bar u < u_g$, whereas for  
the saddle point ($\bar u > u_g$) 
the Gaussian integral is real.

The spectral density is given by
\be
\rho_5(\hz ) &=& \frac 1\pi Z^{f}(\hz)\frac d {d\hz} {\rm Im}Z^{b}(\hz).
\ee
Differentiating the pre-exponential factors gives subleading corrections.
The leading order saddle point result for the spectral density is
thus given by 
\be
\rho_5(\hz) =  \frac 4{-S'''(u_g)}
e^{- \frac 23 (2(\hz_g-\hz))^{3/2}/
\sqrt{-S_b'''(u_g)}}.
\ee
Note that the pre-exponential terms are not determined by a leading
order mean field calculation.
Comparing  this result to  the leading order asymptotic expansion 
of the universal result for the spectral
density at the soft edge
\be
\rho_5(x) = \frac 1\Delta ({\rm Ai}'(x)^2 - x {\rm Ai}^2(x)) \sim \frac{e^{-4x^{3/2}/3}}{\pi x}, \qquad {\rm with} \qquad
x= (z_g-z)/\Delta 
\ee
we find that the two results coincide if we make the identification
\be
\label{defDelta}
\Delta = (-S'''(u_g)/2)^{1/3}.
\ee

\subsubsection{Edge Scaling for $\hz>\hz_g$}

For $\hz > \hz_g$ the leading order saddle point result is determined
by a pair of complex conjugate saddle points. Near the edge of the
spectrum, they are given by Eq. (\ref{saddlepoints}).

In order to  understand the mean field limit for $\hz< \hz_g$ we first discuss the case of $\ha_8 = 0$. 
The saddle points are given by
\be
\sinh \bar r = \pm \frac{i\hz}{\sqrt{\hz^2 -\hm^2}}, \qquad 
\cosh \bar r = \mp \frac{i\hm}{\sqrt{\hz^2 -\hm^2}}.
\ee
When the fermionic and bosonic saddle points have opposite sign, then
the pre-exponential factor vanishes. Therefore this combination of 
the saddle points does not contribute to the partition function. Only one of 
the bosonic saddle points can be reached by deforming the integration contour
and the relevant fermionic saddle point necessarily has the same sign.
This is
the way the supersymmetric method selects the fermionic saddle point and
circumvents the failure of the fermionic replica trick. 

The pre-exponential factors add up to $\sqrt {\hz^2- \hm^2}$ which is canceled
by the contributions from the Gaussian integrals about the saddle point
(up to a factor of $\pi$) resulting in a partition function that is
correctly normalized.

Next we consider the thermodynamic limit at fixed $\ha$. Then the
$1/\sqrt{\hz^2-\hm^2}$ singularity at the edge of the spectrum turns into
a $\sqrt {\hz-\hz_g}$ singularity for which we expect an  Airy like
behavior. The saddle points for $\hz$ close to $\hz_g$ are given
by
\be
\sinh \bar r_\pm = u_g \pm i\left ( \frac
      {2(\hz-\hz_g)}{-S'''(u_g)}\right )^{1/2}, 
\qquad \cosh \bar r = \sqrt{1+\sinh^2 \bar r}.
\ee
The structure of the saddle point approximation to the resolvent is given by
\be
G(z) = P(\bar r_+, \bar r_+) e^{\tilde S_b(\bar r_+) +\tilde S_f(\bar r_+)}
+P(\bar r_+, \bar r_-) e^{\tilde S_b(\bar r_+) +\tilde S_f(\bar r_-)} ,
\label{gmean}
\ee
where the exponents are given in Eq. (\ref{saddles}) and the prefactor
is equal to
\be
(P(s,r) -\frac i2 )\sinh r
\ee
with $ P(s,r) $ defined in Eq. (\ref{pref}).
In the first term in Eq. (\ref{gmean}) the exponents cancel, whereas
in the second term they add up to 
\be
\tilde S_b(\bar r_+) +\tilde S_f(\bar r_-) = i \frac 23 \frac{(2(\hz-\hz_g))^{3/2}}
{\sqrt{-S'''(u_g)}}
\ee
When the saddle points are different the prefactor is suppressed
by $1/\hm $ in agreement with the asymptotic expansion of the universal
expression for the level density near the edge in terms of Airy functions.
We have not worked out the asymptotic behavior for $\hz > \hz_g$ but checked
numerically that the exact expression for the spectral density is 
in agreement with  the expression in terms of Airy function. The
asymptotic domain is already reached when $\hat m \sim  40$ and 
$\hat{a}_8  \sim 1 $.

\newpage

\section{The sign of $W_8$ at small mass}
\label{app:C}

In this appendix we investigate in detail the sign of the partition
function as a function of $W_8$, in the vicinity of
$\hm=0$. Throughout this appendix we will set $\hz=0$.
In particular we can show analytically that at $\hm=0$ the
$N_f=2$ flavor partition function is positive for all values of $\nu$
only if $W_8\geq0$. In contrast the partition function becomes negative for 
odd values of $\nu$ if $W_8<0$. For both signs of $W_8$ we show that the
$N_f=2$ flavor partition function has a local minimum at $\hm=0$: its
first derivative with respect to $\hm$ vanishes, and its second
derivative is positive.

We begin by stating an equivalent integral form of the $N_f=1$ flavor
partition function Eq. (\ref{ZNf1}), by linearizing $\cos^2(\theta)$
through a Gaussian integral and applying an integral representation
for Bessel-$I$ functions:
\be
Z_{N_f=1}^\nu(\hm;\pm\ha_8^2) & = & \exp\left[\pm2\ha_8^2\right]
\int_{-\pi}^\pi \frac{d\theta}{2\pi} \ e^{i\theta\nu} 
\exp\left[{\hm\cos(\theta)\mp4\ha_8^2\cos^2(\theta)}\right]
\nn\\
& = & \exp\left[{\pm2\ha_8^2}\right]
\int_{-\infty}^\infty\frac{dx}{\sqrt{\pi}}\ e^{-x^2}
I_\nu\left(\hm+4\ha_8 x\Big\{{i \atop 1}\Big\} \right).
\label{ZNf1int}
\ee
Here the sign $+(-)$ corresponds to positive (negative)
$W_8$. Following Eq. (\ref{ZNf}) the $N_f=2$ flavor partition function
is then given by 
\be
Z_{N_f=2}^\nu(\hm;\pm\ha_8^2) & = & Z_{N_f=1}^\nu(\hm;\pm\ha_8^2)^2
- Z_{N_f=1}^{\nu-1}(\hm;\pm\ha_8^2) Z_{N_f=1}^{\nu+1}(\hm;\pm\ha_8^2) .
\label{ZNf2}
\ee

At $\hm=0$ the integral in Eq. (\ref{ZNf1int}) 
becomes doable. First, due to the parity of $\mbox{Bessel$-I$}$ 
functions, $I_\nu(-x)=(-)^\nu I_\nu(x)$, 
the zero mass single flavor partition function
vanishes for odd values of $\nu$ (for both signs of $W_8$):
\be
Z_{N_f=1}^{\nu=2k+1}(\hm=0;\pm\ha_8^2) & = & 0.
\ee
At even value of $\nu$ we obtain for a single flavor
\be
Z_{N_f=1}^{\nu=2k}(\hm=0;\pm\ha_8^2) & = & 
\Big\{{i^{2k} \atop  1}\Big\}I_k(2\ha_8^2).
\label{ZNf1nu2k}
\ee
Thus for $N_f=2$ at even $\nu$ we have 
\be
Z_{N_f=2}^{\nu=2k}(\hm=0;\pm\ha_8^2) & = &\Big\{{(-)^{2k} \atop  1}\Big\}
I_k(2\ha_8^2)^2,
\ee
as only the first term in Eq.~(\ref{ZNf2}) is non-vanishing. This is 
obviously real and positive being a complete square,
for both signs of $W_8$. On the other hand when $\nu=2k+1$ the first
term vanishes in Eq.  (\ref{ZNf2}), and we have 
\be
Z_{N_f=2}^{\nu=2k+1}(\hm=0;\pm\ha_8^2) & = &\Big\{{(-)^{2k+1} \atop  1}\Big\}
(-)I_{k}(2\ha_8^2)I_{k+1}(2\ha_8^2). 
\ee
Only for the upper sign $+$ 
the factor $(-)^{2k+1}$ compensates the second minus sign. Knowing
that Bessel-$I$ is a positive function for positive arguments we get
again a positive partition function for $W_8>0$, but a {\it negative}
one for $W_8<0$.

Next we investigate whether or not the point $\hm=0$ is a relative
extremum. Using the following Bessel identity,
\be
I_\nu^\prime(x) &=& \frac12 (I_{\nu-1}(x)+I_{\nu+1}(x))
\ee
it is easy to see that 
\be
\frac{\partial}{\partial\hm}
Z_{2}^\nu(\hm;\pm\ha_8^2)\Big|_{\hm=0} & = & \frac12\Big(
Z_{1}^\nu(0;\pm\ha_8^2)
( Z_{1}^{\nu-1}(0;\pm\ha_8^2)+Z_{1}^{\nu+1}(0;\pm\ha_8^2))\\
&&-Z_{1}^{\nu-2}(0;\pm\ha_8^2)Z_{1}^{\nu+1}(0;\pm\ha_8^2)
-Z_{1}^{\nu-1}(0;\pm\ha_8^2)Z_{1}^{\nu+2}(0;\pm\ha_8^2)
\Big).\nn
\label{ZNf2I}
\ee
Because of its parity this expression vanishes for both even and odd
$\nu$, independent of the sign of $W_8$. 

To see if this is a local minimum we compute the second derivative,
given by 
\be
\frac{\partial^2}{\partial\hm^2}
Z_{2}^\nu(\hm;\pm\ha_8^2)\Big|_{\hm=0} & = & \frac14\Big(
2Z_{1}^\nu(0;\pm\ha_8^2)^2
-2Z_{1}^{\nu-2}(0;\pm\ha_8^2)Z_{1}^{\nu+2}(0;\pm\ha_8^2)
\nn\\
&&+Z_{1}^{\nu-1}(0;\pm\ha_8^2)^2
-Z_{1}^{\nu-3}(0;\pm\ha_8^2)Z_{1}^{\nu+1}(0;\pm\ha_8^2)\nn\\
&&+Z_{1}^{\nu+1}(0;\pm\ha_8^2)^2
-Z_{1}^{\nu-1}(0;\pm\ha_8^2)Z_{1}^{\nu+3}(0;\pm\ha_8^2)
\Big).
\label{ZNf2II}
\ee
For even $\nu$ only the first line is non-vanishing due to parity, and we have 
\be
\frac{\partial^2}{\partial\hm^2}
Z_{2}^{\nu=2k}(\hm;\pm\ha_8)\Big|_{\hm=0} & = & \frac12
\Big\{{(-)^{2k} \atop  1}\Big\}
\Big(I_{k}(2\ha_8^2)^2-I_{k-1}(2\ha_8^2)I_{k+1}(2\ha_8^2)\Big).
\ee
This is positive for both signs of $W_8$:
\be
I_{k}(x)^2-I_{k-1}(x)I_{k+1}(x)= 2\int_0^1dt\,t\ 
I_{k}(xt)^2\ >\ 0.
\label{ZLS}
\ee
Taking equation Eq. (\ref{ZNf2II}) for odd $\nu$, the first line
vanishes due to parity and the second and third line do contribute to give:
\be
\frac{\partial^2}{\partial\hm^2}
Z_{2}^{\nu=2k+1}(\hm;\pm\ha_8)\Big|_{\hm=0} & = & 
\frac14\Big\{{(-)^{2k-2} \atop  1}\Big\}
\Big(I_{k}(2\ha_8^2)^2-I_{k-1}(2\ha_8^2)I_{k+1}(2\ha_8^2)\Big)\nn\\
&&
+\frac14\Big\{{(-)^{2k+2} \atop  1}\Big\}
\Big(I_{k+1}(2\ha_8^2)^2-I_{k}(2\ha_8^2)I_{k+2}(2\ha_8^2)\Big)\ .
\ee
This is once more positive for both signs of $W_8$, because of the
identity (\ref{ZLS}).
In conclusion the second derivative is always positive and we have a
relative minimum for small $\hm$.


\end{document}